\documentclass[11pt,a4paper]{article}
\usepackage{citesort}
\usepackage{amsmath}
\usepackage{graphicx}%
\usepackage{amsfonts}%
\usepackage{amssymb}
\usepackage{epsfig}

\newlength{\dinwidth}
\newlength{\dinmargin}
\newcommand{\vl}{\pmb{l}}
\newcommand{\vk}{\pmb{k}}
\newcommand{\vp}{\pmb{p}}
\newcommand{\vq}{\pmb{q}}
\newcommand{\ve}{\pmb{\eta}}
\newcommand{\vP}{\pmb{P}}
\newcommand{\vu}{\pmb{u}}
\newcommand{\vr}{\pmb{r}}
\newcommand{\vR}{\pmb{R}}

\newcommand{\bea}{\begin{eqnarray}}
\newcommand{\eea}{\end{eqnarray}}

\newcommand{\prm}{\sigma}
\newcommand{\beq}{\begin{equation}}
\newcommand{\eeq}{\end{equation}}
\newcommand{\beqar}[1]{\begin{eqnarray}\label{#1}}
\newcommand{\eeqar}{\end{eqnarray}}

\newcommand{\beeq}{\begin{eqnarray}}
\newcommand{\eeeq}{\end{eqnarray}}

\def\Ba{D}

\begin{document}

\title{{~}\\[1cm]
{\Large \bf 
Baryon scattering at high energies:\\
wave function, impact factor, and gluon radiation
 }}
\author{ 
{~}\\
J.~Bartels$\,{}^{a)}\,$\thanks{Email: bartels@mail.desy.de} 
\hspace{1ex} and \;
L.~Motyka$\,{}^{a),b)}\,$\thanks{E-mail: motyka@th.if.uj.edu.pl}
\\[10mm]
{\it\normalsize ${}^{a)}$ II Institute for  Theoretical Physics, 
University of Hamburg, Germany}\\
{\it\normalsize $^{b)}$ Institute of Physics, Jagellonian University,
Krak\'{o}w, Poland} \\}

\date{November 14, 2007}

\maketitle
\thispagestyle{empty}

\begin{abstract}
\noindent
The scattering of a baryon consisting of three massive quarks is investigated 
in the high energy limit of perturbative QCD. A model of a relativistic 
proton-like wave function, dependent on valence quark longitudinal and 
transverse momenta and on quark helicities, is proposed, and we derive the 
baryon impact factors for two, three and four $t$-channel gluons. We find that 
the baryonic impact factor can be written as a sum of three pieces: 
in the first one a subsystem consisting of two of the three quarks behaves 
very much like the quark-antiquark pair in $\gamma^*$~scattering, 
whereas the third quark acts as a spectator.
The second term belongs to the odderon, whereas in the third 
($C$-even) piece all three quarks participate in the scattering. 
This term is new and has no analogue in $\gamma^*$~scattering. 
We also study the small~$x$ evolution of gluon radiation for each of these 
three terms. The first term follows the same pattern of gluon radiation as the 
$\gamma^*$--initiated quark-antiquark dipole, and, in particular, it contains
the BFKL evolution followed by the $2\to 4$ transition vertex 
(triple Pomeron vertex). The odderon-term is described by the standard 
BKP evolution, and the baryon 
couples to both known odderon solutions, the Janik-Wosiek solution
and the BLV solution. Finally, the $t$-channel evolution of the third term 
starts with a three reggeized gluon state which then, via a new
$3\to 4$ transition vertex, couples to the four gluon (two-Pomeron) state. 
We briefly discuss a few consequences of these findings, in particular 
the pattern of unitarization of high energy baryon scattering amplitudes.
\end{abstract}

\begin{flushright}
\vspace{-23.0cm}
{DESY 07--198}\\
hep-ph/0711.2196\\
\end{flushright}
\thispagestyle{empty}

\newpage

\section{Introduction}

In recent years deep inelastic electron proton or electron nucleus 
scattering (DIS) at small $x$ has attracted much interest, and it has 
stimulated intense studies of high energy QCD. At high energies, 
the total cross section of a virtual photon scattering on a  
target, in a first approximation, can be described in terms of a 
photon impact factor and a Balitsky-Fadin-Kuraev-Lipatov (BFKL) Green's 
function~\cite{bfkl,bfklsum,bfklnl}. 
When restricting to the large $N_c$ limit, and assuming a large target, 
unitarity corrections to this first approximation are described by 
the nonlinear Balitsky-Kovchegov (BK) equation~\cite{balitsky,kov1} which, 
in the language of BFKL Green's functions, represents the infinite sum of 
fan diagrams ~\cite{blv1}.  
The BK equation was initially obtained in the $s$-channel color dipole 
picture (in the large $N_c$ limit)~\cite{niko,mueller}. 
Beyond the large $N_c$ limit one has to include the full color structure 
of the $2\to 4$ reggeized gluon vertex~\cite{jbar} which leads
to the Balitsky hierarchy of integral equations~\cite{balitsky} or 
to the Jalilian-Marian--Iancu--McLerran--Weigert--Leonidov--Kovner
(JIMWLK) equations~\cite{jimwalk}.
In many of these calculations the incoming virtual photon plays a 
vital r\^{o}le: its large virtuality $Q^2$ justifies the use of 
perturbation theory, and its impact factor consists of a 
quark-antiquark pair which forms a color dipole configuration. 
This simple structure is also intimately connected with the 
fan-like structure of the diagrams resumed by the nonlinear BK equation.

The advent of the LHC challenges us with the task of developing a theoretical 
understanding of scattering in high energy proton-proton 
collisions, which is related to the structure of unitarity corrections 
in baryon-baryon scattering. In this paper we will perform a study of the 
high energy behavior of baryon scattering within perturbative QCD.   
It is clear that the problem of
high energy nucleon scattering is much more complex than it was in the 
virtual photon case. First of all, in nucleon-nucleon or nucleon-nucleus 
scattering the incoming projectiles are nonperturbative, and the accuracy
of perturbative calculations is not under good theoretical control. 
We shall circumvent this problem by studying a fictitious scattering process
of a heavy and small {\em baryonium} system, in analogy to 
the heavy {\em onium} proposed as a test case for perturbative unitarity 
corrections in DIS~\cite{mueller}. 
For such processes the perturbative calculations provide reliable 
results. Next, the baryonium scattering is expected to differ 
significantly from the onium scattering. The main reason is the difference
of the color structure: in contrast to the color dipole the baryon is a color 
singlet formed by three valence quarks. Also, the application of the 
large $N_c$~limit which played the crucial r\^{o}le in the construction 
of the dipole model is rather difficult in the baryon case: one needs
exactly $N_c$ quarks to build the color singlet of the $SU(N_c)$ group,
and this system becomes rather complex for $N_c \to \infty$.
In fact, a few years ago, it was explicitly pointed out~\cite{praszalowicz} 
that the simple picture of gluon radiation which has emerged in the QCD 
dipole picture does not work in the case of an incoming three quark 
color singlet system; however, no alternative solution had been derived.
Thus, we shall address the issue of gluon radiation from three quarks
at $N_c = 3$, within a perturbative baryonic system and compare with 
the perturbative quark-antiquark system. 

The basic and universal object that characterizes properties of the baryon 
is its wave function. Inspired by the success of the concept of the 
{\em photon wave function} \cite{niko} which turned out to be very 
fruitful in studies of high energy scattering, we start from a local 
three-fermion quark current operator with the quantum numbers of the proton 
and construct a relativistic invariant infinite momentum frame 
wave function for the lowest Fock component of the baryon, 
consisting of three valence quarks. The resulting wave function 
contains a non-trivial dependence on quark helicities and angular momenta.
For the current operator we chose the baryonic operator
proposed by Ioffe~\cite{ioffe}, which has been shown to provide a reasonable 
phenomenological prescription of the nucleon properties~\cite{vbraun1}.
In order to take into account the nonperturbative nature of the baryon 
we make use of the Borel transform technique which has been developed in the 
context of QCD sum rules.
     
This paper is not intended yet to deal with a detailed phenomenology of the 
baryon structure and scattering, --- thus we do not attempt, for example,  
to tune the obtained wave function to describe the existing data on 
proton form-factors and high energy scattering. Nevertheless, apart from 
developing a theoretical laboratory for studying scattering of baryon states 
at high energies, one may hope that our perturbative analysis finds 
{\em structures} which remain also relevant beyond the perturbatively safe 
region. An extrapolation of our results on the heavy baryonium 
to the realistic proton case may, therefore, very well allow for some useful 
phenomenology. More detailed studies in this direction will be left for 
future work.

Starting from integrals over squares of these baryonic wave functions 
and coupling $t$-channel gluons to the three quark lines we define 
baryonic impact factors, in close analogy with the photon impact factor 
in deep inelastic electron proton scattering. The small-$x$ evolution 
of baryon scattering amplitude will be analyzed, again,
following the strategy  developed in the context of the virtual 
photon scattering~\cite{jbar,bw,be}. First we consider, in lowest order, the 
elastic scattering of the baryonic system on a single quark: by coupling two 
$t$-channel gluons to the three-quark system, the baryonic 
impact factor is obtained. Three or four $t$-channel 
gluons appear if one considers, again at lowest order, multi-particle
amplitudes, e.g. $3 \to 3$ processes in a suitably defined high energy limit.  
In the next step, one considers higher order diagrams in the leading 
logarithmic approximation: this leads to rapidity evolution equations,
describing the radiation of gluons from the three-quark system.

Our main results are the following. We propose a model of the 
baryon wave function with a non-trivial quark helicity and angular 
momentum structure. Then we express the baryon impact factor in terms of
the wave function, for an arbitrary number of coupling gluons. 
The obtained baryonic impact factor can be written as a sum of several pieces, 
each of them having its own evolution equation.
First, there is a term in which one pair out of the three quarks scatters
whereas the third quark acts as a spectator. Although the two quarks  
which participate in the interaction are in a color anti-triplet 
configuration, they behave very much like the quark--antiquark pair in the 
photon case. In the lowest order, two $t$-channel gluons couple to this 
quark pair. In higher order the two gluons start to reggeize and to 
produce the full BFKL ladder, while the third quark of the baryon state 
remains an inactive spectator. Also, the well-known $2 \to 4$ gluon vertex 
appears, indicating the beginning of the same fan-like structure as 
in the quark--antiquark case. Altogether, this piece of the 
baryon impact factor radiates gluons in very much the same way as the 
quark-antiquark pair in the photon case.    

Next, there is the odderon term, similar to the one discussed 
in ~\cite{ewerz_odd}: here all three quarks participate, and the $t$ channel 
state carries $C=-$. In lowest order, three gluons couple to the three 
quarks; in higher order the state evolves according to the 
Bartels-Kwieci\'{n}ski-Prasza\l{}owicz (BKP) 
evolution equation~\cite{bkp_bar,bkp_kp}.                    

Finally, a third, $C$-even, piece of the baryonic impact factor appears 
in which again all three quarks participate. This piece has no counterpart 
in the quark-antiquark case and, together with the odderon, 
it makes the baryon really behaving differently from the photon 
(or the vector meson). The state consists of one reggeized gluon with
even signature and two usual odd reggeized gluons. It obeys the BKP evolution
in the three Reggeon channel and it decays into four reggeized gluons 
via a new gauge invariant $3\to 4$ reggeized gluon vertex.

The paper is organized as follows. We begin with a short section describing 
the general framework in which our calculations are carried out. We then 
(Section~3) turn to the baryon wave function which enters the baryon impact 
factor. In the following Section~4 we describe the baryon impact factor and 
its decomposition into the three pieces described above, and in Section~5 
we discuss the rapidity evolution of these pieces. Section~6 contains a short 
discussion of the baryonic impact factor in configuration space, 
and in Section~7 we analyze the quark--diquark limit of the baryon
wave function. Finally, in Section~8 we summarize our results and discuss a 
few potential implications.

\section{The framework}

In our calculation we will follow the analysis of the scattering of a virtual 
photon described in ~\cite{bw,be}. 
In leading order the scattering of a virtual photon  
off a quark is described by the exchange of two gluons. The coupling to the photon 
is described by the photon impact factor, $D_{2;0}$, which most easily is obtained by the 
energy discontinuity of a closed quark loop (Fig.~1). Making use of the Regge factorization, the 
same impact factor can also be used in other elastic scattering processes, e.g. in the
scattering of a virtual photon on a heavy onium target. Higher order corrections, in the 
leading logarithmic approximation, lead to the reggeization of the $t$-channel gluons and to 
the exchange of a BFKL Pomeron between the photon impact factor and the target.    

\begin{figure}
\begin{center}
\epsfig{file=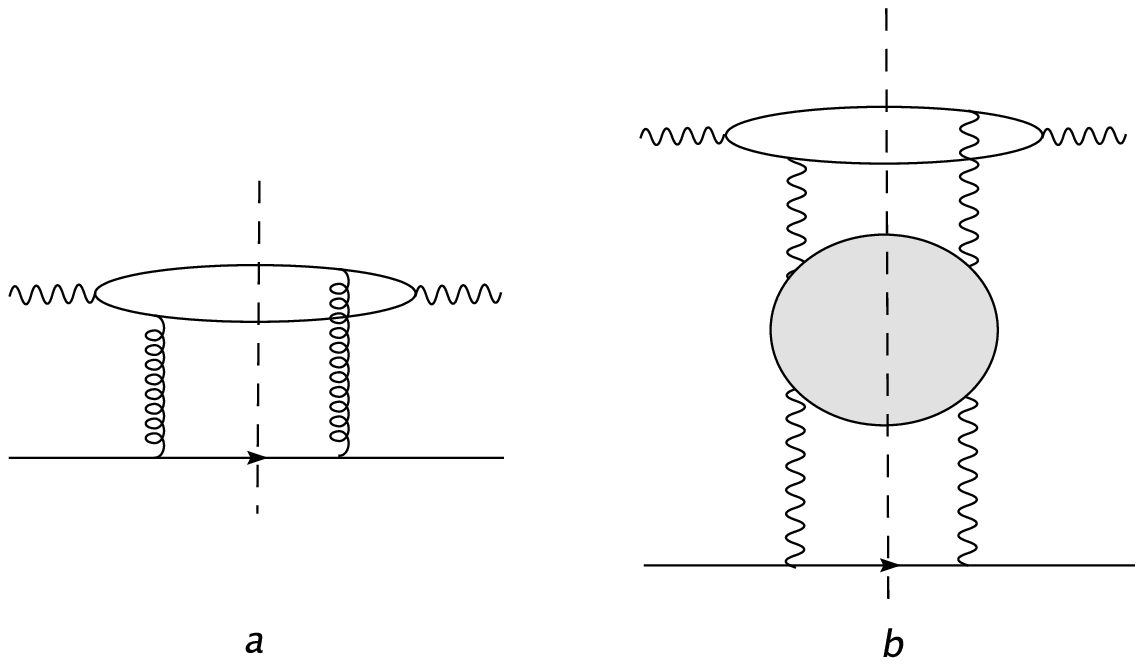,width=12cm,height=6cm}\\
\end{center}
\caption{\it Energy discontinuity of the $2\to2$ process: $\gamma^* +q \to \gamma^* +q$.}
\end{figure}

If one is looking for corrections containing more than two reggeized $t$-channel gluons  
one has to go beyond the leading logarithmic approximation. In the elastic scattering process 
$\;\gamma^* + q \to \gamma^* + q\;$, both leading order and NLO corrections retain the structure 
of a single ladder. A $t$-channel state with four reggeized gluons appears first in NNLO.
A convenient way to avoid the complications connected with such a high order calculation 
is the study of multi-particle processes, e.g the $3 \to 3$ process $\;\gamma^* + q +q \to 
\gamma^* + q +q\;$, the scattering of a virtual photon on two independent quarks (Fig.~2) 
in the triple Regge limit. This process depends upon three independent energy variables, and 
the triple energy discontinuity can be easily computed in the approximation where, in each order 
perturbation theory, the maximal number of large energy logarithms is kept. The lowest order 
contribution is described by the exchange of four gluons. In higher order, these $t$-channel 
gluons reggeize and start to interact. As discussed in detail in ~\cite{bw,be}, the all-order 
result can be cast into the two sets of diagrams shown in Fig.~3.

\begin{figure}
\begin{center}
\epsfig{file=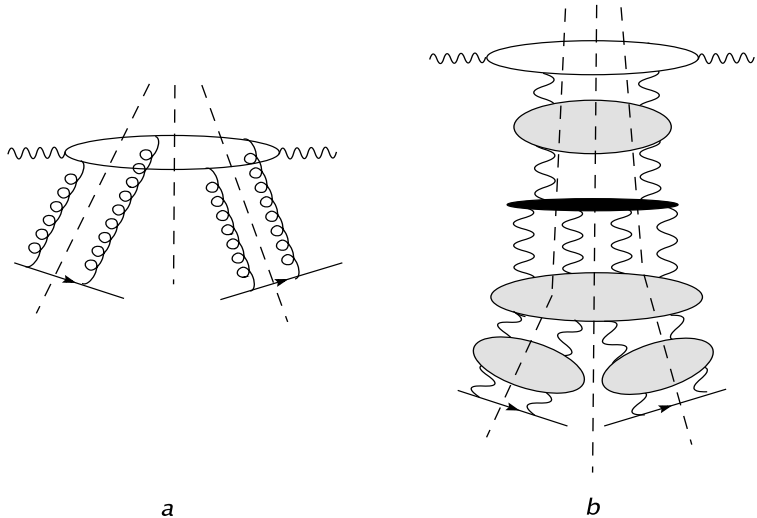,width=10cm,height=7cm}\\
\caption{\it Multiple energy discontinuities of the $3\to 3$ process: 
$\gamma^* +q+q \to \gamma^* +q+q$: \hspace{4cm}
\centerline{(a)~lowest order diagram, (b) two examples of higher order 
diagrams.}
}
\end{center}
\end{figure}

\begin{figure}
\begin{center}
\epsfig{file=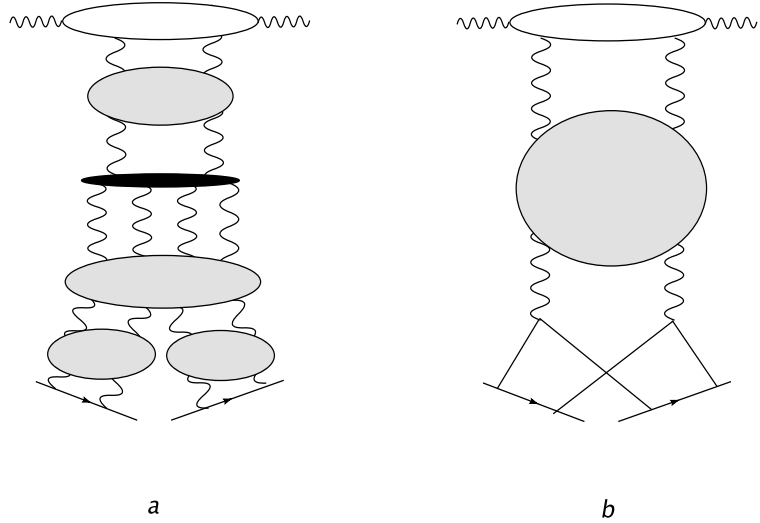,width=10cm,height=7cm}\\ 
\end{center}
\caption{\it Decomposition of the sum of all diagrams in Fig.~2b into 
(a) irreducible and (b) reggeizing pieces.} 
\vspace*{8mm}
\end{figure}

The first term starts, at the photon impact factor, with a BFKL Green's 
function, then undergoes the transition into the four gluons and continues 
with the BKP evolution of the four gluon state. 
In the large-$N_c$ limit, the four gluon state turns into two noninteracting 
BFKL systems, i.e.\ we see the beginning of the fan-diagram structure 
of the BK equation. The second term consists of a simple BFKL Green's 
function, with higher order splittings of the reggeized gluons at the lower 
end. As a remarkable feature of this results, in both contributions only 
two reggeized gluons couple to the photon impact factor, despite the fact 
that diagrams with four gluons --- such as the one shown in Fig.~2a --- 
are included: the apparent `disappearance' of these contributions is a 
result of the gluon reggeization which manifests itself in generalized 
bootstrap relations. 

The same strategy can be used to investigate $t$-channel states with higher number of 
$t$-channel reggeized gluons. For example, six gluons appear in the $8$-point amplitude
$\;\gamma^* + q +q+q  \to \gamma^* + q +q +q\;$, i.e. the scattering of a virtual photon 
on three independent quarks. The analysis of this case has been investigated in
~\cite{be}.  

Although these results are --- initially --- derived in the context of a higher order 
multi-particle processes (e.g. the $3 \to 3$ scattering process), 
they nevertheless can be used also in a $2 \to 2$ process. The diagrams shown in Fig.~3 
satisfy Reggeon unitarity equations in all three $t$-channels. Taking the discontinuity 
across the four Reggeon state, the partial wave above this can be used to construct 
the four Reggeon state in the $2 \to 2$ process shown in Fig.~4.          

\begin{figure}
\begin{center}
\epsfig{file=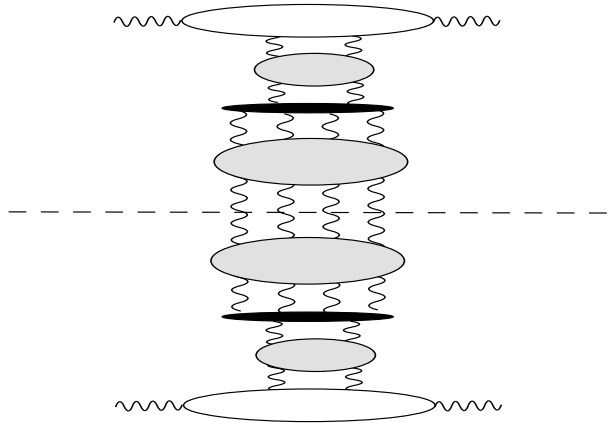,width=8cm,height=5.6cm}\\
\end{center}
\caption{\it Four gluon contribution to the Reggeon unitarity equation of
elastic $\gamma^*\gamma^*$ scattering.} 
\end{figure}
     
In this paper we will apply the same construction, replacing the virtual photon 
by a three quark system. Modeled by the four fermion operator introduced 
by Ioffe in the context of the QCD sum rules~\cite{ioffe,svz}, 
the incoming `baryon' splits into three quarks which then couple to $2$, $3$, 
or $4$~gluons. In order to take into account the non-local  nature of the 
incoming baryonic bound state we introduce a form factor: we employ 
a technique used in the QCD sum rules~\cite{svz} and use the Borel 
transform of the perturbative expression~\cite{lip_bal,iv_kir}. 
The exponential nature of this form factor also guarantees 
the convergence of the momentum integrals inside the impact factor.

\section{The baryon wave function}

\begin{figure}[t]
\centerline{\epsfig{file=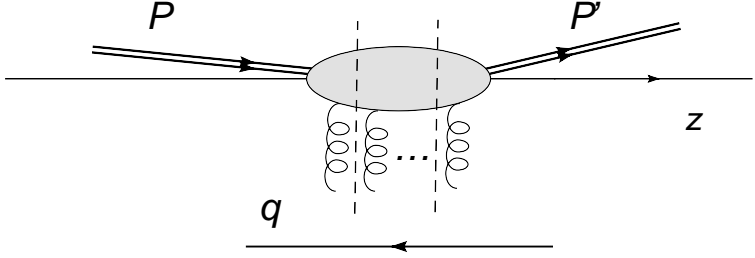,width=10cm}}
\caption{\it Multiple discontinuity of the impact factor for  
elastic baryon scattering.}
\label{fig:kinematics}
\end{figure}

We consider the multiple discontinuity of a non-forward baryon impact factor 
in elastic high energy scattering. Large momenta are directed along the $z$-axis,
and the incoming and outgoing baryons move at small angles with respect to 
the $z$-axis, as shown in Fig.~\ref{fig:kinematics}. Their momenta $P$, $P'$ 
have a large ``$+$'' light-cone component, $P^+$, and their transverse momenta  
are denoted by $\vP$, $\vP'$, respectively.
We introduce the light-like vector $q^\mu = (q,0,0,-q)$, $q^2 = 0$  
with $\,s \,=\, (P+q)^2 \simeq 2P\cdot q\,$, and we assume that $s$ is large:
$\;s\gg M^2,\,\vP^2,\, {\vP'} ^2$. 
The quark momenta $p_i$ are
\beq 
p_i^\mu \;= \; (p_i^0,\vp_i,p_i^z),\quad  p_i^+ \; = \; p_i^0 +p_i^z,
\quad p_i^- = p_i^0 - p_i^z. 
\eeq
For the longitudinal quark momenta it will sometimes be convenient 
to use the notation
\beq
p_{i}^+ \,= \,\alpha_{i} P^+, \quad p_{i}^- \,=\, \beta_{i} q^- .
\eeq
We shall use $\;\hat p = \gamma_\mu p^\mu = \gamma \cdot p\;$ for contraction 
of four-vectors and Dirac $\gamma$~matrices. 
The adopted model of the proton state is defined by
\beq
\langle 0 | \eta(0) \, | N(P,\lambda) \, \rangle \,=\, 
A_N \; w_\lambda (P), 
\eeq
where $w_\lambda (P)$ is the proton spinor with momentum $P$ and helicity 
$\lambda$, 
\beq
\eta(x) = 
{\varepsilon\,}_{\kappa_1 \kappa_2 \kappa_3 }\;
[(u^{\kappa_1} (x))^T C\gamma^\mu\, u^{\kappa_2}(x)] \; \gamma_\mu \gamma_5 \, 
d^{\,\kappa_3}(x)
\eeq
is the baryonic Ioffe current ~\cite{ioffe}, $C$ is the charge conjugation 
matrix, and $\kappa_i$ are color indices. 
The Ioffe operator is not the only possible choice of the baryon current, ---
in the context of distribution amplitudes, the possible baryonic operators 
for the proton were classified in Ref.\ \cite{vbraun2}, and it was shown that
the Ioffe current gives a rather good description of baryon 
form-factors~\cite{vbraun1}. We therefore chose, as a test case, 
the Ioffe operator to model the baryonic impact factor.\footnote{It is worthwhile to stress 
that our baryon wave functions are different from the distribution
amplitudes. In the collinear approach one probes the baryon with a hard 
external scale, $Q^2$, and the baryon structure is represented by series 
of distribution amplitudes with increasing twist, that is with increasing power-like 
suppression at large $Q^2$. The distribution amplitudes depend on 
the quark longitudinal momenta, and they obey evolution
equations in $\log Q^2$. In contrast to that, we are interested in
the baryon wave function with full momentum dependence probed at a moderate
momentum scale, and the evolution applies to the rapidity of gluons radiated from the 
baryon impact factor.}
\begin{figure}[t]
\centerline{\epsfig{file=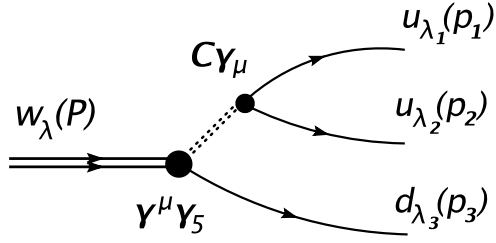,width=6.5cm}}
\caption{\it The proton vertex as given by the Ioffe current.}
\label{fig:ioffe}
\end{figure}

For the calculation of the baryonic impact factor we will need the 
matrix elements (Fig.~\ref{fig:ioffe}) in the helicity basis,
\beq
\left[ \, \bar d _{\lambda_3}(p_3) \, \gamma_5 \gamma_\mu\, w_{\lambda} (P)\, 
\right] 
\,\cdot \, 
\left[ \, \bar u _{\lambda_1}(p_1) \, \gamma^\mu\, 
C\gamma^0\, u^*_{\lambda_2}(p_2) \, 
\right]. 
\eeq
In the second term we can also write:
\beq
\left[ \, \bar u _{\lambda_1}(p_1) \, \gamma^\mu\, v_{\lambda_2}(p_2) \, \right]
\eeq 
where $v$ (in the Dirac notation) denotes the $v$-spinor of the $u$ quark.

\subsection{The massless quark case}

Using the calculus described by Brodsky and Lepage~\cite{bro_lep} we 
compute the Dirac spinor matrix elements. The details of the calculations 
are described in Appendix A. For simplicity, we start from the massless 
quark case, and the case of massive quarks will be analyzed afterwards.
Thus we obtain:  
\beq
\label{spinorplus}
{\left[ \, \bar d _{\lambda}(p_3) \, \gamma_5 \gamma_\mu\, w_{\lambda}(P) \, \right] 
\,\cdot \, 
\left[ \, \bar u _{\lambda_1}(p_1) \, \gamma^\mu\, C\gamma^0\, 
u^*_{\lambda_2}(p_2) \, \right] \over
\sqrt{\alpha_1\alpha_2\alpha_3}}
\; = \;  \rule{0mm}{10mm}
\eeq
\[
 = \; 2\,\lambda\;\delta_{-\lambda_1,\, \lambda_2} \;  
 \left\{ \;
\delta_{\lambda_1,\, \lambda} \; \left[
\left( {\vp_2 \over \alpha_2} - \vP\right)\,\cdot\,
\left( {\vp_1 \over \alpha_1} - {\vp_3 \over \alpha_3} \right)\; - \; 
i\lambda \; \left({\vp_2 \over \alpha_2} - \vP \right)\,\times\,
\left( {\vp_1 \over \alpha_1} - {\vp_3 \over \alpha_3} \right)\; \right]\right.
\; + \; \nonumber
\rule{0mm}{10mm}
\]
\[
\left.
+ \; \delta_{\lambda_2,\, \lambda} \; \left[
\left({\vp_1 \over \alpha_1} - \vP\right)\,\cdot\,
\left( {\vp_2 \over \alpha_2} - {\vp_3 \over \alpha_3} \right)\; - \; 
i\lambda \; \left( {\vp_1 \over \alpha_1}-\vP\right)\,\times\,
\left( {\vp_2 \over \alpha_2} - {\vp_3 \over \alpha_3} \right)\; \right]
\right\}, 
\nonumber
\rule{0mm}{10mm}
\]
and,
\beq
\label{spinorminus}
{\left[ \, \bar d _{-\lambda}(p_3) \, \gamma_5\gamma_\mu\, w_{\lambda}(P) \, \right] 
\,\cdot \, 
\left[ \, \bar u _{\lambda_1}(p_1) \, \gamma^\mu\, C\gamma^0\, u^*_{\lambda_2}(p_2)  \, \right] \over
\sqrt{\alpha_1\alpha_2\alpha_3}} \;  =  \;  \rule{0mm}{10mm}
\eeq
\[
= \; 2M\;\,\delta_{-\lambda_1,\, \lambda_2} \;
\left\{ \;
\delta_{\lambda_1,\, \lambda} \;
\ve_{\lambda}\, \cdot\,
\left( {\vp_2 \over \alpha_2} - {\vp_3 \over \alpha_3} \right) \right.
\; + \; 
\left. 
\, \delta_{\lambda_2,\, \lambda} 
\; \ve_{\lambda}\, \cdot\,
\left( {\vp_1 \over \alpha_1} - {\vp_3 \over \alpha_3} \right)\; \right\}, 
\rule{0mm}{10mm}
\vspace{5mm}
\]
where the transverse complex vector $\ve_{\lambda}$ is defined by
\beq
 \ve_{\lambda} = (1,i\lambda), \qquad \lambda = \pm 1,
\eeq
and $\vP = \vp_1 +\vp_2 + \vp_3$ is the transverse momentum of the incoming 
baryon. The cross product of two transverse vectors 
$\vp_1 = (p_1^x,p_1^y)$ and $\vp_2 = (p_2^x,p_2^y)$ 
should be  understood as a number
$\;\vp_1\times\vp_2 \; = \; p_1 ^x \, p_2 ^y \,-\,  p_1 ^y\, p_2 ^x.\,$
It turns out that formula (\ref{spinorplus}) may be re-expressed in a more 
compact form, by using the vectors $\ve_{\lambda}$ with the following identity:
\beq
\label{etaprod}
(\vp_1 \cdot \ve_{\lambda})\; (\vp_2 \cdot \ve_{-\lambda}) \; = \;
(p_1^x\, + \,i\lambda p_1^y)\,(p_2^x\, - \,i\lambda p_2^y) \; = \;
p_1^x\, p_2^x \, + p_1^y\, p_2^y \, + 
\, i \lambda ( p_1^y\, p_2^x \, - \,  p_1^x\, p_2^y\,) \; = \; 
\vp_1 \cdot \vp_2 \; - \; i \,\lambda \, \vp_1 \times \vp_2,
\eeq
which holds for any pair of transverse vectors, $\vp_1$ and $\vp_2$. 
Using this relation one gets:
\[
{\left[ \, \bar d _{\lambda}(p_3) \, \gamma_5 \gamma_\mu\, w_{\lambda}(P) \, \right] 
\,\cdot \, 
\left[ \, \bar u _{\lambda_1}(p_1) \, \gamma^\mu\, C\gamma^0\, 
u^*_{\lambda_2}(p_2) \, \right] \over
\sqrt{\alpha_1\alpha_2\alpha_3}}
\; = \;  \rule{0mm}{10mm}
\]
\[
 = \;\;  2\,\delta_{-\lambda_1,\, \lambda_2} \;\; \times \;
\left\{ \;
\delta_{\lambda_1,\, \lambda} \;
\left[\, \ve_{\lambda}\, \cdot  
\left( {\vp_2 \over \alpha_2} - \vP\right)\,
\right]
\left[\, \ve_{-\lambda}\, \cdot \, 
\left( {\vp_1 \over \alpha_1} - {\vp_3 \over \alpha_3} \right)
\right]
\right.
\; + \; \nonumber\\ 
\]
\beq
\left.
\delta_{\lambda_2,\, \lambda} \;
\left[ \, \ve_{\lambda}\, \cdot  
\left( {\vp_1 \over \alpha_1} - \vP\right)\,
\right]
\left[\, \ve_{-\lambda}\, \cdot \, 
\left( {\vp_2 \over \alpha_2} - {\vp_3 \over \alpha_3} \right)
\right]
\right\}.
\vspace{5mm}
\label{spinorplusb}
\eeq
In what follows, we shall express all formulae in this compact notation.

\begin{figure}[t]
\leavevmode
\centerline{
\begin{tabular}{lll}
\epsfig{file=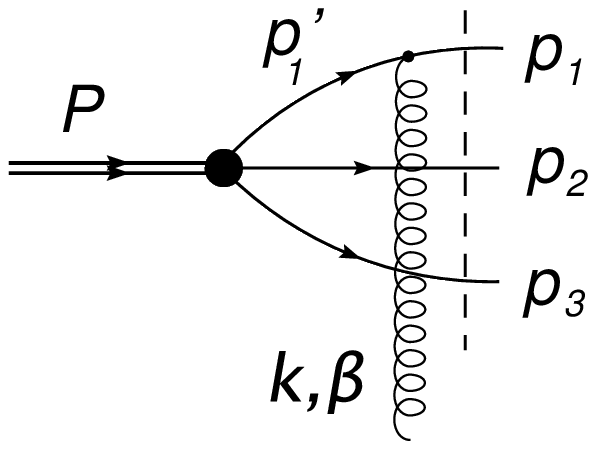,width=4.5cm} &
\epsfig{file=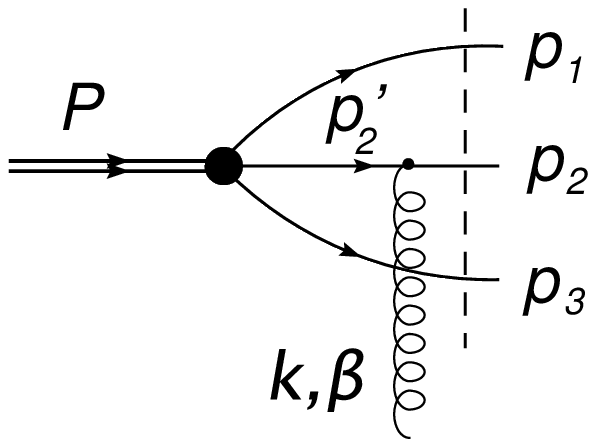,width=4.5cm} &
\epsfig{file=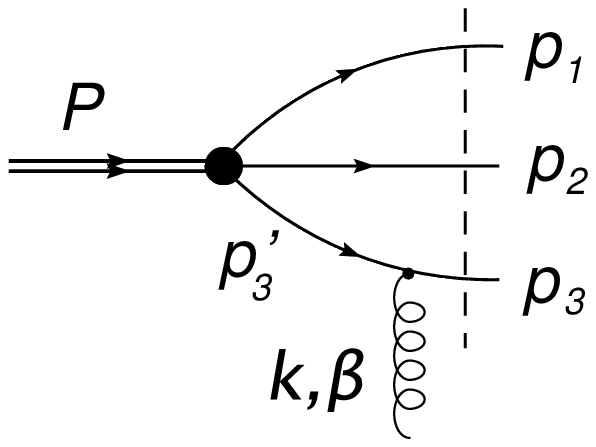,width=4.5cm} \\
{\Large\bf a)} & {\Large\bf b)} & {\Large\bf c)} \\
\end{tabular}
}
\caption{\it Single gluon coupling to the baryon $\to X$
transition.
\label{fig:theta}
}
\end{figure}

Next we couple a gluon of momentum $k = \beta q + \vk$ to one of the quark 
lines with momentum $p_i$ (Fig.~\ref{fig:theta}). 
Fixing the momenta of the outgoing quarks at $p_1$, $p_2$, and $p_3$, the 
quark line to the left of the gluon vertex carries momentum $p_i-k$. 
Using, at the gluon vertex, the eikonal approximation, one arrives at  
the spinorial factor $\hat{q}$. With $\;2\, p_i\cdot q \;=\; \alpha_i\, s \; 
\gg \; 
\vp_i^2,\vk^2,\; $etc.\ , one obtains, for the upper $u$ quark,
\beq
\label{step1}
\bar u(p_1)\, \hat q \,(\hat{p_1}-\hat{k}) \; = \; 
{2 p_1\cdot q}\; \bar u(p_1-k) \, + \, \ldots\,, 
\eeq
where $\,\ldots\,$ stands for terms which are power suppressed in $s$.
An analogous expression holds for the $d$ quark, whereas for the second 
$u$ quark we use:
\beq
\label{step2}
(\hat{p_2}- \hat{k}) \,\hat{q}\, u^*(p_2)\; = \; 
2\,p_2 \cdot q \, u^*(p_2-k)\,+\, \ldots\,. 
\eeq
As a result, on the r.h.s.\ of Eqs.\ (\ref{step1}) and (\ref{step2}),
the transverse momentum of the quark 
spinor coincides with the transverse momentum of the internal quark line next 
to the baryon vertex. The sum of the outgoing transverse momenta equals
\beq
\vp_1+\vp_2+\vp_3 = \vP +\vk.
\eeq

Matrix elements corresponding to multi-gluon couplings to spinor lines
may be simplified 
by iterating Eq.\ (\ref{step1}) in the following way: 
\beq
\label{multishift}
\bar u(p)\; \hat{q} \; 
[\gamma\, \cdot \, (p- k_{1})]\; \hat q \; \ldots \;
\hat q \; [\gamma\, \cdot \, (p- k_{1}- \ldots -k_n)] \;
 \; \simeq \; (2 p\cdot q)^n \, \bar u(p-k_1-\ldots -k_n).
\eeq
For completeness, we remind that, in the case of an outgoing antiquark,
an additional minus sign appears:
\beq
\label{step3}
- (\hat{p}- \hat{k})\, \hat q\, v(p) \; = \; 
{-2 p\cdot q}\; v(p-k) \, + \, \ldots ,
\eeq
This minus sign is due to the opposite direction of the momentum along the 
antifermion line. Similarly:
\beq
[-\gamma\, \cdot \, (p - k_1- \ldots -k_n)\,]\; \hat q \ldots  
\hat{q} [-\gamma\, \cdot \, (p- k_{1})] \;
\hat q \; v(p) \simeq \; (-2 p\cdot q)^n \, v(p-k_1-\dots -k_n).
\eeq
This change in sign plays a crucial r\^{o}le in the photon 
impact factor~\cite{niko}. 

We are now ready to describe the amplitude for the process:
baryon + gluon $\to$ 3 quarks, defined by the diagrams shown in 
Fig.~7. We define the shifted momentum of the upper quark
\beq
\vp_1' = \vp_1 - \vk,
\eeq 
with 
\beq
\vp_1'+\vp_2+\vp_3 = \vP,
\eeq
and use 
\beq
{p_1'}^2 = \alpha_1 \left( M^2 +\vP^2 - 
\frac{{\vp_1'}^2}{\alpha_1}- \frac{{\vp_2}^2}{\alpha_2}-\frac{{\vp_3}^2}
{\alpha_3} \right), 
\eeq
(and analogous expressions for the gluon coupling to quark lines $2$ and $3$).
We introduce the amplitudes
\beeq
\Theta_{\;\lambda} ^{(\lambda_1,\lambda_2)\, \lambda} 
(\{\alpha_{i}\},\{\vp_1,\vp_2,\vp_3\},\vP)
& = & \lambda\,{\cal N}_{\Theta}\;\,
{2\,\sqrt{\alpha_1\alpha_2\alpha_3}
\over M^2 + \vP^2 
- {\vp_1^2 \over \alpha_1} 
- {\vp_2^2 \over \alpha_2}
- {\vp_3^2 \over \alpha_3} 
} \; \;
\delta_{-\lambda_1,\, \lambda_2}\;\;\delta^{(2)} (\vp_1+\vp_2+\vp_3 - \vP) \,
\, \times
\nonumber 
\\
& \times &  \rule{0mm}{10mm}
\left\{ \;
\delta_{\lambda_1,\, \lambda} \;
\left[\, \ve_{\lambda}\, \cdot  
\left( {\vp_2 \over \alpha_2} - \vP\right)\,
\right]
\left[\, \ve_{-\lambda}\, \cdot \, 
\left( {\vp_1 \over \alpha_1} - {\vp_3 \over \alpha_3} \right)
\right]
\right.
\; + \; \nonumber\\ 
 & &  \rule{0mm}{10mm}
\left.
\delta_{\lambda_2,\, \lambda} \;
\left[ \, \ve_{\lambda}\, \cdot  
\left( {\vp_1 \over \alpha_1} - \vP\right)\,
\right]
\left[\, \ve_{-\lambda}\, \cdot \, 
\left( {\vp_2 \over \alpha_2} - {\vp_3 \over \alpha_3} \right)
\right]
\right\}, 
\label{Theta1}
\\
\Theta_{\;\lambda} ^{(\lambda_1,\lambda_2)\, -\lambda} 
(\{\alpha_{i}\},\{\vp_1,\vp_2,\vp_3\}
;\,\vP)
& = & \rule{0mm}{10mm}
{\cal N}_{\Theta}\;\, {2M\,\sqrt{\alpha_1\alpha_2\alpha_3}
\over M^2 + \vP^2 
- {\vp_1^2 \over \alpha_1} 
- {\vp_2^2 \over \alpha_2}
- {\vp_3^2 \over \alpha_3} 
} \; \;
\delta_{-\lambda_1,\, \lambda_2}\;\;\delta^{(2)} (\vp_1+\vp_2+\vp_3 - \vP) \,
\, \times
\nonumber
\\
& \times & \rule{0mm}{10mm}
\left\{ \;
\delta_{\lambda_1,\, \lambda} \;
\ve_{\lambda}\, \cdot\,
\left( {\vp_2 \over \alpha_2} - {\vp_3 \over \alpha_3} \right) \right.
\; + \; 
\, \delta_{\lambda_2,\, \lambda} 
\left. \; \ve_{\lambda}\, \cdot\,
\left( {\vp_1 \over \alpha_1} - {\vp_3 \over \alpha_3} \right)\; \right\}. 
 \label{Theta2}
\eeeq
Here the upper three indices of $\Theta$ denote the helicities of the upper two
u quarks with momenta $\vp_1$ and $\vp_2$, and the lower d quark with 
momentum $\vp_3$, respectively. The subscript refers to the 
helicity $\lambda$ of the incoming baryon. We leave the normalization 
constant $\;{\cal N}_{\Theta}\;$ unspecified here; 
the normalization will be fixed at the level of baryon wave function.
The amplitudes for the diagrams shown in Figs.~7a--7c are then 
simply obtained from (\ref{Theta1}), (\ref{Theta2})  
by the replacements $\vp_1 \to \vp_1'$, $\vp_2 ,\to \vp_2'$ 
and $\vp_3 \to \vp_3'$, respectively.\footnote{To be precise, the functions
$\Theta$ give the {\em  momentum dependent part} of the scattering amplitudes, 
up to a global normalization factor, that is proportional to the 
strong coupling constant $g$. Obviously, the color factors are not 
accounted for in (\ref{Theta1}) and (\ref{Theta2}), --- they will be treated 
explicitly later on.}

Note that, for each of the three diagrams, the denominator is just  
the energy denominator in non-covariant perturbation theory, for instance
one obtains for Fig.~7a,
\beq
E_{\mathrm{baryon}} - E_{\,\mathrm{3\; quark}} \; = \;
\frac{1}{P^+} \left( M^2 + \vP^2 -
\frac{{\vp_1'}^2}{\alpha_1}- \frac{{\vp_2}^2}{\alpha_2}-\frac{{\vp_3}^2}
{\alpha_3} \right). 
\eeq
The amplitudes should be invariant under Lorentz boosts in the transverse
directions, parametrized by a four-velocity 
$u^\mu \simeq (1,\vu,0)$, $|\vu| \ll 1$: 
\beq
\vp_{i} \;\to\; \vp'_{i} \; \simeq \; \vp_{i} \, + \,{1\over 2}\,p_{i} ^+ \,\vu,
\qquad {p}_{i} ^+ \; \to \;  {p'}_{i} ^+ \; \simeq \; {p}_{i} ^+ .
\eeq
The numerators are manifestly invariant under these transformations, 
and the denominators may be also rewritten in an explicitly invariant form 
using the identity
\beq
-\vP^2 + {\vp_1 ^2 \over \alpha_1} +
{\vp_2 ^2 \over \alpha_2} +
{\vp_3 ^2 \over \alpha_3}
 \;\; = \;\;
\alpha_1\alpha_2\left({\vp_1 \over \alpha_1} - {\vp_2 \over \alpha_2}\right)^2 
\;+\;
\alpha_1\alpha_3\left({\vp_3 \over \alpha_3} - {\vp_1 \over \alpha_1}\right)^2
\;+\;
\alpha_2\alpha_3\left({\vp_2 \over \alpha_2} - {\vp_3 \over \alpha_3}\right)^2,
\eeq
or
\beq
-\vP^2 + {\vp_1 ^2 \over \alpha_1} +
{\vp_2 ^2 \over \alpha_2} +
{\vp_3 ^2 \over \alpha_3}
 \; = \;
{(\vp_1 - \alpha_1 \vP)^2 \over \alpha_1} 
\; + \;  {(\vp_2 - \alpha_2 \vP)^2 \over \alpha_2} 
\; + \;  {(\vp_3 - \alpha_3 \vP)^2 \over \alpha_3}. 
\eeq

The denominators have poles for the invariant mass of the three-quark system
equal to the proton transverse mass. 
Clearly, this is a consequence of using a point-like
vertex for the proton-quark coupling and neglecting the bound state effects.
These effects cannot be described within perturbative QCD and should 
be modeled. Following ~\cite{svz,lip_bal} we propose a model that preserves Lorentz and helicity structures of the perturbative expressions, where the bound state effects are absorbed into the Borel transform.

The Borel transform of a function $f(s)$ is defined in the standard way:
\beq
{\cal B}_s[\,f\,](M_B^2) \; = \; 
\lim_{n \to \infty} \\  
{s^{n+1} \over n!}\left( -{d\over ds} \right)^n\; f(s), \qquad 
s\to \infty, \quad s/n \to M_B^2,
\eeq
where $M_B$ is the Borel parameter. 
In order to model the baryon scattering amplitude we shall apply two independent Borel 
transforms w.r.t.\ the negative virtualities: $Q^2=-P^2$ of the incoming and 
$Q'^2=-P'^2$ of the outgoing baryon,
to the perturbative amplitudes obtained with the point-like vertex. 
Formulae (\ref{Theta1}) and (\ref{Theta2}) were presented for $P^2 = M^2$. 
The corresponding 
formulae for general virtualities are obtained by substitutions 
$M^2 \to P^2$ in the 
denominators. In the baryon impact factor, the virtuality $P^2$ 
appears only in the energy denominator of the vertex amplitude 
$\Theta_{\;\lambda} ^{(\lambda_1,\lambda_2)\, \lambda_3} 
(\{\alpha_{i}\},\{\vp_i\};\,\vP)$ of the incoming baryon, 
and the virtuality $P'^2$ only in the denominator of the  amplitude  
$\left[\Theta_{\;\lambda} ^{(\lambda_1,\lambda_2)\, \lambda_3} 
(\{\alpha_{i}\},\{\vp_i\};\,\vP')\right]^*$ of the outgoing state 
(see Sec.~4 for more details).
Therefore the two Borel transforms may be performed independently for 
each $\Theta$, that is already at the level of the baryon wave function.
Thus we evaluate
\beq
{\cal B}_{Q^2}\left[{1 \over Q^2+M_X ^2}\right](M_B^2) \; = \; 
\exp\left(-M_X^2 / M_B^2\right).
\eeq
This result, applied to the amplitudes $\Theta$, leads to the substitution
\beq
{1 \over P^2 + \vP^2 - \sum_{i=1} ^3 {\vp_i^2 \over \alpha_i}} 
\; \longrightarrow \;  
-\exp \left[ -{1 \over M_B^2} \left( 
\sum_{i=1} ^3 {\vp_i^2 \over \alpha_i} \, - \, \vP^2\right) \right]. 
\eeq

Before we complete the model we shall perform some simplifications.
We shall absorb into the wave functions a phase space factor 
$(\alpha_1\alpha_2\alpha_3)^{-1}$ that appears in the baryon impact 
factor as a result of on-mass-shell conditions of the cut quark lines.
In this way, the factor $\,\sqrt{\alpha_1\alpha_2\alpha_3}\,$ 
present in the amplitudes $\Theta$ will be removed from the wave functions.
Obviously, the integration measure will be suitably modified as well.
For simplicity, we introduce a normalization constant, ${\cal N}$, 
of the wave function that will be fixed later. 
Thus, we choose the natural value of the Borel parameter $M_B = M$ and 
obtain a model of the baryon wave function,
\beeq
\Psi_{\;\lambda} ^{(\lambda_1,\lambda_2)\, \lambda} 
(\{\alpha_{i}\},\{\vp_{i}\};\,\vP) \; 
& = &
\lambda \; 
{\cal N} e^{ -\frac{1}{M^2} \left(-\vP^2 +
\frac{{\vp_1}^2}{\alpha_1}+ \frac{{\vp_2}^2}{\alpha_2}+\frac{{\vp_3}^2}
{\alpha_3} \right) } 
\; \;
\delta_{-\lambda_1,\, \lambda_2}\;\;\delta^{(2)} (\vp_1+\vp_2+\vp_3 - \vP) 
\; \times \nonumber \\
& \times &  \rule{0mm}{10mm}
\left\{ \;
\delta_{\lambda_1,\, \lambda} \;
\left[\, \ve_{\lambda}\, \cdot  
\left( {\vp_2 \over \alpha_2} - \vP\right)\,
\right]
\left[\, \ve_{-\lambda}\, \cdot \, 
\left( {\vp_1 \over \alpha_1} - {\vp_3 \over \alpha_3} \right)
\right]
\right.
\; + \; \nonumber\\ 
 & + &  \rule{0mm}{10mm}
\left.
\delta_{\lambda_2,\, \lambda} \;
\left[ \, \ve_{\lambda}\, \cdot  
\left( {\vp_1 \over \alpha_1} - \vP\right)\,
\right]
\left[\, \ve_{-\lambda}\, \cdot \, 
\left( {\vp_2 \over \alpha_2} - {\vp_3 \over \alpha_3} \right)
\right]
\right\}, 
\label{psiplus}
\\
\Psi_{\;\lambda} ^{(\lambda_1,\lambda_2)\, -\lambda} 
(\{\alpha_{i}\},\{\vp_{i}\}
;\,\vP) 
& = & \rule{0mm}{10mm}
{\cal N} e^{ -\frac{1}{M^2} \left(- \vP^2 +
\frac{{\vp_1}^2}{\alpha_1} +\frac{{\vp_2}^2}{\alpha_2} + \frac{{\vp_3}^2}
{\alpha_3} \right) }
\; \;
\delta_{-\lambda_1,\, \lambda_2}\;\;\delta^{(2)} (\vp_1+\vp_2+\vp_3 - \vP) 
\;\times
\nonumber \\
& \times & \rule{0mm}{10mm} \;\;\;\;
M\; \left\{ \;
\delta_{\lambda_1,\, \lambda} \;
\ve_{\lambda}\, \cdot\,
\left( {\vp_2 \over \alpha_2} - {\vp_3 \over \alpha_3} \right) \right.
\; + \; 
\, \delta_{\lambda_2,\, \lambda} 
\left. \; \ve_{\lambda}\, \cdot\,
\left( {\vp_1 \over \alpha_1} - {\vp_3 \over \alpha_3} \right)\; \right\}. 
\label{psiminus}
\eeeq
Clearly, the functions $\Psi$ given by Eqs. 
(\ref{psiplus}) and (\ref{psiminus}) are symmetric under the interchange 
of the $u$~quarks, labeled by 1 and~2. When combined with the 
anti-symmetry in the color degrees of freedom it implies that 
the full wave function is anti-symmetric under interchange
of the $u$~quarks, as it must be. Interestingly enough, a similar Gaussian 
form of the wave function was proposed long ago~\cite{bl2} and it was shown 
to provide a good description of the nucleon form-factor 
data~\cite{bol_kro,mdiehl}. An important difference of our model, however, 
is the presence of angular momenta of the quarks. The baryon angular momentum 
structure following from the model is most transparent in the coordinate 
representation and will be discussed is Section~6.

The above derivation of the baryon wave function is based on perturbative QCD methods
combined with the Borel transform technique. Clearly, we are not able to control the accuracy
of this procedure for the proton as it is a genuine non-perturbative object. Therefore the
obtained wave functions can be only considered as a theoretically inspired model of the
proton wave function. Therefore, in the next part we give the formulae for the wave function of a baryon consisting of three quarks with the same mass $m$, coming in two different flavors. 
These formulae will permit to consider the fictitious case of a large quark mass, for which the 
baryon becomes heavy and small, and the perturbative computation of its wave function 
and scattering is formally justified.

\subsection{Massive quarks}

We now apply the procedure described in the previous section
to the case of the massive quarks. We skip the details of the derivation 
and present the result for the helicity amplitudes $\Theta$ of the transition: 
baryon to quarks, in which all three quarks were assumed to have the 
mass $m$:\vspace{2mm}
\beq 
\Theta_{\;\lambda} ^{(\lambda_1,\lambda_2)\, \lambda_3} 
(\{\alpha_{i}\},\{\vp_i\},\vP)
\; = \; {\cal N}_{\Theta}\;
{2\,\sqrt{\alpha_1\alpha_2\alpha_3}
\over M^2 + \vP^2 
- {\vp_1^2 + m^2\over \alpha_1} 
- {\vp_2^2 + m^2 \over \alpha_2}
- {\vp_3^2 + m^2 \over \alpha_3} 
} \; \;\delta^{(2)} (\vp_1+\vp_2+\vp_3 - \vP) \,
\, \times 
\nonumber
\eeq
\beeq
& \times &  \rule{0mm}{10mm} 
\left\{
\delta_{\lambda,\, \lambda_1} \,
\delta_{\lambda,\, \lambda_2} \,
\delta_{\lambda,\, \lambda_3} \; 
m\,{\alpha_1+\alpha_2 \over \alpha_1\alpha_2}\;   
\left( {\vp_3 \over \alpha_3} \,-\,
{\vp_1 + \vp_2 \over \alpha_1+\alpha_2}
\right)\cdot \ve_{-\lambda}
\; + \; \right.
\nonumber \\
& + & \rule{0mm}{10mm} 
\lambda \, 
\delta_{\lambda,\, \lambda_1} \,
\delta_{\lambda,\, \lambda_2} \,
\delta_{\lambda,\, -\lambda_3} \;\; 
m\,{\alpha_1+\alpha_2 \over \alpha_1\alpha_2}\;  
\left( M - {m\over \alpha_3} \right) \; +
\nonumber \\
& + & \rule{0mm}{10mm} 
\delta_{\lambda,\,   -\lambda_1} \,
\delta_{\lambda,\,   -\lambda_2} \,
\delta_{\lambda,\, \lambda_3} \; 
m\,{\alpha_1+\alpha_2 \over \alpha_1\alpha_2}\;   
\left({\vp_1 + \vp_2 \over \alpha_1+\alpha_2} \, - \, \vP
\right)\cdot\ve_{\lambda}
\; + \;
\nonumber \\
& + & \rule{0mm}{10mm} 
\lambda\; 
\delta_{\lambda,\, \lambda_1}\,
\delta_{\lambda,\, -\lambda_2}\,
\delta_{\lambda,\, \lambda_3} \;
\left[
\left({\vp_2 \over \alpha_2}-\vP\right)\cdot\ve_{\lambda}
\;
\left({\vp_1 \over \alpha_1}-{\vp_3 \over \alpha_3} \right) \cdot
\ve_{-\lambda}
\; + \;
m\, \left({M \over \alpha_3} \, - \, {m\over \alpha_1\alpha_2}\right)
\right] \; + \; 
\nonumber \\
& + & \rule{0mm}{10mm} 
\lambda\;
\delta_{\lambda,\, -\lambda_1}\,
\delta_{\lambda,\, \lambda_2}\,
\delta_{\lambda,\, \lambda_3} \;
\left[
\left({\vp_1 \over \alpha_1}-\vP\right)\cdot\ve_{\lambda}
\;
\left({\vp_2 \over \alpha_2}-{\vp_3 \over \alpha_3} \right)\cdot
\ve_{-\lambda}
\; + \;
m \left(
{M \over \alpha_3} \, - \, {m \over \alpha_1\alpha_2}\right)
\right] \; + \;
\nonumber \\
& + & \rule{0mm}{10mm} 
\delta_{\lambda,\,  \lambda_1}\,
\delta_{\lambda,\,  -\lambda_2}\,
\delta_{\lambda,\, -\lambda_3} \; 
\left[M\,
\left( {\vp_2 \over \alpha_2} - {\vp_3 \over \alpha_3} \right)
\, \cdot\, \ve_{\lambda} \; 
+ \;\, m \;{1-\alpha_3 \over \alpha_3}\; 
\left( {\vp_2 \over \alpha_2} \,-\, {\vp_1 + \vp_2 \over \alpha_1 + \alpha_2}
\right) \cdot\ve_{\lambda}\, 
\right] 
\; + \;  
\nonumber \\
& + & \rule{0mm}{10mm} 
\left.
\delta_{\lambda,\,  -\lambda_1}\,
\delta_{\lambda,\,  \lambda_2}\,
\delta_{\lambda,\,-\lambda_3} \; 
\left[M\,
\left( {\vp_1 \over \alpha_1} - {\vp_3 \over \alpha_3} \right)
\cdot\ve_{\lambda} \; 
+ \;\, m \;{1-\alpha_3 \over \alpha_3}\; 
\left( {\vp_1 \over \alpha_1} \,-\, {\vp_1 + \vp_2 \over \alpha_1 + \alpha_2}
\right) \cdot\ve_{\lambda}\,
\right] \right\}.
\label{Thetamass}
\eeeq
Note, that using relation (\ref{etaprod}) and taking $m\to 0$ one easily 
recovers formulae (\ref{Theta1}) and~(\ref{Theta2}). 
The above formulae are promoted to the baryon wave functions $\Psi$, 
by going through the same steps as in the massless case, i.e.
using the Borel transform and absorbing the $\alpha$ factors into the 
phase space factor. The final expressions for the
wave functions are obtained from Eq.~(\ref{Thetamass}) by the replacement 
which combines both steps:
\beq
{\cal N}_{\Theta}\;\,
{2\,\sqrt{\alpha_1\alpha_2\alpha_3}
\over M^2 + \vP^2 
-\sum_{i=1} ^3 {\vp_i^2 + m^2\over \alpha_i}} 
\;\; \longrightarrow \;\; 
{\cal N}\;
\exp\,\left[ 
\,-{1\over M^2} \,
\left(
\sum_{i=1} ^3 {\vp_i^2 + m^2\over \alpha_i} 
\, - \, \vP^2  \,\right)\right]. 
\vspace{3mm}
\eeq

\section{Baryon impact factors}

\subsection{General structure}

The amplitudes $\Psi$ may be combined with their complex conjugates
to obtain the baryon impact factor. For the case of two gluons 
coupled to lines $3$ and $1$ we illustrate one example in Fig.~8.
It was shown in the previous section that, in the high energy limit 
the spinorial part of the multiple discontinuity can be expressed in terms 
of universal matrix elements given by Eqs.\ (\ref{spinorplus}) 
and (\ref{spinorminus}), where the momenta of the quarks are evaluated 
at the quark-proton vertex. Also the denominator is determined by the 
virtuality of the quark to which the first gluon couples, and  
the  virtuality can expressed in terms of the momenta of the quarks at the 
proton vertex. Thus, the impact factor can be obtained from overlap integrals,
i.e. products of wave functions with suitably adjusted momenta.
As an example, we specify the overlap function corresponding Fig.~8:

\begin{figure}
\begin{center}
\epsfig{file=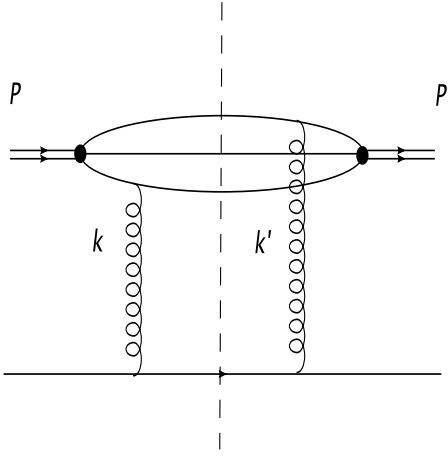,width=6cm,height=6cm}\\
\end{center}
\caption{\it A contribution to the two gluon exchange in baryon-quark scattering.}
\end{figure}

\beeq
{\cal F}^{\;\lambda \lambda'}(\{\vk,\vk'\}; \vP,\vP') \; = \; 
\sum_{\lambda_1,\lambda_2,\lambda_3} \;
\int [d^2 \vp_{i}] \, [d\alpha_{i}] \;\; 
\left[
\Psi_{\;\lambda'} ^{(\lambda_1,\lambda_2)\,\lambda_3}
\left(\{\alpha_{i}\},\{\vp_1',\vp_2,\vp_3\};\vP'\right)   
\right]^* 
\nonumber
\\  
\times \;
\Psi_{\;\lambda} ^{(\lambda_1,\lambda_2)\,\lambda_3}
\left(\{\alpha_{i}\},\{\vp_1,\vp_2,\vp_3'\};\vP\right),
\eeeq
with the integration measure
\beq
[d^2\vp_{i}] \; = \;  d^2\vp_1 \; d^2\vp_2\; d^2\vp_3\, , \qquad
[d\alpha_{i}] \; = \;  
{d\alpha_1}\,
{d\alpha_2}\,
{d\alpha_3}\;\;
\delta(\alpha_1+\alpha_2+\alpha_3-1).
\eeq  
In the overlap functions ${\cal F}^{\;\lambda \lambda'}$,
the upper helicity labels refer to the incoming and outgoing baryon states, 
respectively. Analogous overlap functions are defined for the other gluon 
couplings, and for the full impact factor we will have to sum over all 
diagrams. When evaluating the sum over the intermediate helicities 
$\lambda_1$, $\lambda_2$, and $\lambda_3$ and summing over all diagrams, 
one finds, for the forward direction $\vP=\vP'=0$, helicity conservation, 
i.e.\ the impact factor vanishes for $\lambda = - \lambda'$.

Before including the remaining energy integrals and the color factors we generalize to the
case of $3$ and $4$ $t$-channel gluons. As outlined in Section 2, we have to consider multiple 
energy discontinuities. An example is shown in Fig.~9.
\begin{figure}
\begin{center}
\epsfig{file=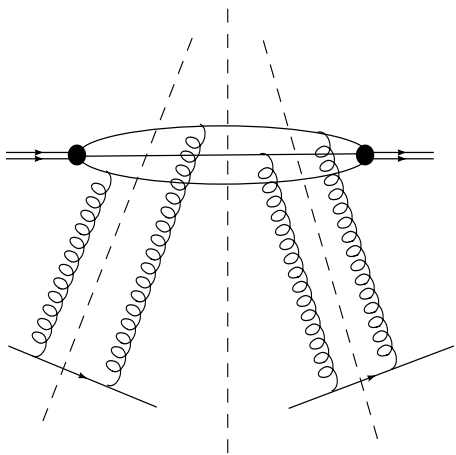,width=6cm,height=6cm}\\
\end{center}
\caption{\it A contribution to the four gluon exchange in the scattering of a 
baryon on two independent quarks.}
\end{figure}
For one of the discontinuity lines (in the case of Fig.~9, the central line) 
we fix the intermediate quark momenta and denote them by 
$\vp_1,\; \vp_2\; ,\vp_3$. 
The corresponding overlap function (Fig.~10) takes the form 
\begin{eqnarray}
{\cal F}^{\;\lambda \lambda'}(\{\vk_i,\vk'_j\}; \vP,\vP') \;=\; 
\sum_{\lambda_1,\lambda_2,\lambda_3} \;
\int [d^2 \vp_{i}] \, [d\alpha_{i}] \;\; 
\left[
\Psi_{\;\lambda'} ^{(\lambda_1,\lambda_2)\,\lambda_3}
\left(\{\alpha_{i}\},\{\vp_1+\vk_1',\vp_2 +\vk_2',\vp_3+\vk_3'\};\vP'\right)   
\right]^*
\nonumber \\ 
\times \;
\Psi_{\;\lambda} ^{(\lambda_1,\lambda_2)\,\lambda_3}
\left(\{\alpha_{i}\},\{\vp_1 - \vk_1,\vp_2 - \vk_2,\vp_3 -\vk_3\};\vP\right) 
,
\label{overlap1}
\end{eqnarray} 
where $\vk_i$ ($\vk'_j$) is the sum of momenta delivered by the gluons to the
quark line~$i$ to the left (right) of the central cutting line. Obviously,
\beq
\sum_i \vk_i +\sum_j \vk'_j =\vP' - \vP.
\eeq

\begin{figure}
\begin{center}
\epsfig{file=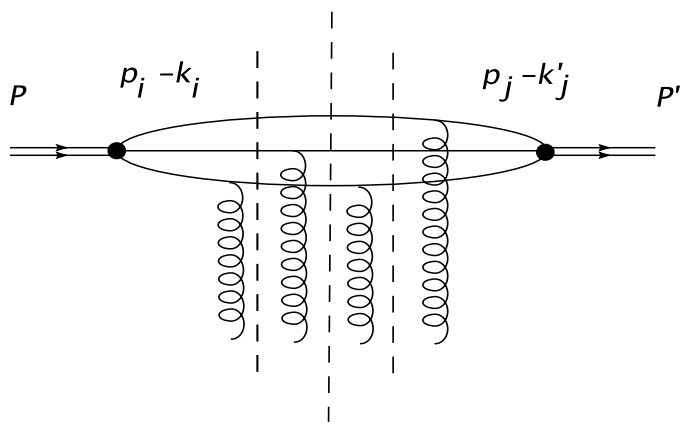,width=9cm,height=5.5cm}\\
\end{center}
\caption{\it Example of a diagram defining the overlap function for a multiple 
discontinuity in Fig.~9.}
\end{figure}   
Since, as a consequence of the exponential form factor, all transverse 
momentum integrals are finite, we are allowed to shift, for each diagram separately, 
the loop momenta, such that to the right of the incoming baryon vertex, 
the momenta become $\vp_1$, $\vp_2$, and $\vp_3$. At the outgoing baryon vertex the 
momenta are $\vp_1+\vl_1$, $\vp_2+\vl_2$, and $\vp_3+\vl_3$, where 
$\vl_i$ is the sum of transverse momenta of {\it all} gluons coupled to the 
quark line $i$ (Fig.~10):
\beq
\vl_{i} \; = \; \sum_{j \in L_{i}}\, (\,\vk_j + \vk_j'\,).
\eeq
In the general case the overlap function can be written as:
\beq
{\cal F}^{\;\lambda \lambda'}(\{\vl_{i}\}; \vP,\vP') \;=\; 
\sum_{\lambda_1,\lambda_2,\lambda_3} \;
\int [d^2 \vp_{i}] \, [d\alpha_{i}] \;\; 
\left[
\Psi_{\;\lambda'} ^{(\lambda_1,\lambda_2)\,\lambda_3}
\left(\{\alpha_{i}\},\{\vp_{i}+\vl_{i}\};\vP'\right)   
\right]^* \;
\Psi_{\;\lambda} ^{(\lambda_1,\lambda_2)\,\lambda_3}
\left(\{\alpha_{i}\},\{\vp_{i}\};\vP\right).
\label{overlap2}
\eeq

\begin{figure}[t]
\centerline{\epsfig{file=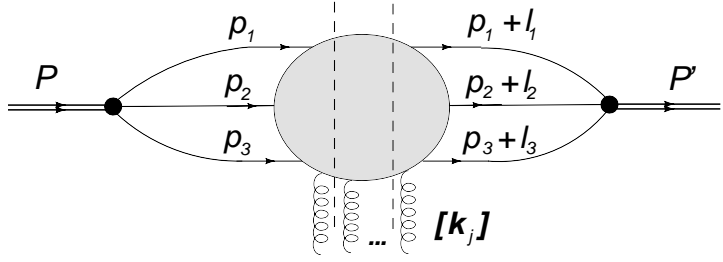,width=9.5cm}}
\caption{\it Baryon impact factor.} 
\end{figure}

We now complete the definition of the baryonic impact factor (see Fig.~11).
The general impact factor $\,{\cal I}^{(N)} _{AB}\,$ for the transition 
$\,A\to B\,$, with $N$ gluons being coupled in the eikonal approximation,   
is defined as
\beq
{\cal I}_{AB} ^{(N)} \; = \; \int 
{d\beta_1 \over 2\pi} \, 
{d\beta_2 \over 2\pi} \,
\ldots \,
{d\beta_{N-1} \over 2\pi}\;\, 
{q_{\mu_1}\,q_{\mu_2}\,\ldots\, q_{\mu_{N}} \over s}\;\;
\mathrm{Disc}\,_{N-1}\;{\cal S}_{AB} ^{\,\mu_1\,\mu_2\,\ldots\,\mu_N},
\eeq
where ${\cal S}_{AB} ^{\,\mu_1\,\mu_2\,\ldots\,\mu_N}\,$ 
represents the amputated transition amplitude.
In particular, for the elastic scattering of a single quark inside the 
baryon one obtains:
\beq
{\cal I}_{qq}^{(N)} \; = \; {1 \over N_c}\;\, \mathrm{Tr}\,
[\,t^{a_N} t^{a_{N-1}} \ldots t^{a_1}\,] \;\, I_{qq} ^{(N)}, 
\eeq
where
\beq
 I_{qq} ^{(N)} \; = \; (-ig)^N \, .
\eeq
The baryon impact factor for $N$ $t$-channel gluons is then given by
\beq
\label{bcal}
{\cal B}_N ^{\,\lambda\lambda'} (\{\vl_{i}\}; \vP,\vP') \; = \;
I_{qq} ^{(N)} \; \sum_{\mathrm{diagrams}}
{\cal F}^{\;\lambda \lambda'}(\{\vl_{i}\}; \vP,\vP') \;\;
{\cal C}_N(\mathrm{diagram}),
\eeq
where the color factor reads:
\beq
{\cal C}_N(\mathrm{diagram}) \; = \;
{\varepsilon^{\kappa'_1\kappa'_2\kappa'_3} \,
\varepsilon^{\kappa_1\kappa_2\kappa_3} \over 3!}
\left[\,t^{a_l} t^{a_{l-1}} \ldots t^{a_1} \right]_{\kappa'_1\kappa_1}\;
\left[\,t^{b_m} t^{b_{m-1}} \ldots t^{b_1} \right]_{\kappa'_2\kappa_2}\;
\left[\,t^{c_n} t^{c_{n-1}} \ldots t^{c_1} \right]_{\kappa'_3\kappa_3}.
\label{colortrace}
\eeq
In (\ref{bcal}) the sum extends over all diagrams, and 
the numbers $l$, $m$ and $n$~of gluons, that couple to quark lines 1, 2 and~3, 
take all possible values between $1$ and $N$ with the constraint $l+m+n=N$.
The overlap function 
$\;{\cal F}^{\;\lambda \lambda'}(\{\vl_{i}\}; \vP,\vP') \;$ is evaluated
for each diagram separately, and in each case it contains a global delta 
function of the transverse momenta:
\beq
{\cal F}^{\;\lambda \lambda'}(\{\vl,\vl'\}; \vP,\vP') = 
{F}^{\;\lambda \lambda'}(\vl_1,\vl_2,\vl_3)\;\;
\delta^{(2)}\left(\sum_{i} \vl_{i} + \vP - \vP'\right).
\label{F-function}
\eeq
We impose the normalization condition:
\beq
\label{fnorm}
\left.
{F}^{\;\lambda \lambda'}(\vl_1,\vl_2,\vl_3)\,
\right| _{\;\vl_1 = \vl_2 = \vl_3 = 0} \; = \; \delta^{\lambda \lambda'}. 
\eeq
Correspondingly, we also extract a delta function from the impact factor:
\beq
{\cal B}_N^{\,\lambda\lambda'} (\{\vl_{i}\}; \vP,\vP') \; 
= \;B_N ^{\,\lambda\lambda'} (\{\vl_{i}\}) \;
\delta^{(2)}\left(\sum_{i} \vl_{i} + \vP - \vP'\right).
\eeq

In the following we will restrict ourselves to the forward direction,
$\vP=\vP'=0$. Because of helicity conservation for the impact factor, 
we always have $\lambda=\lambda'$, and we will drop the upper helicity labels, i.e.\ 
${F}^{\;\lambda \lambda'}(\vl_1,\vl_2,\vl_3) \to {F}(\vl_1,\vl_2,\vl_3)$ etc.
We will go through the cases of $N=2$, $N=3$, 
and $N=4$ gluons. We therefore define, for fixed $N$, the functions
$B_{N;0\,}(\{\vl_{i}\})$ projected on the $C$-even channel through
\beq
\left. {\cal B}_N ^{\,\lambda\lambda} (\{\vl_{i}\}; \vP,\vP') 
\,\right|_{\vP=\vP'=0} ^{{\mathrm C-even}}
\; = \;
B_{N;0\,}(\{\vl_{i}\}) \;\;\;\delta^{(2)}\left( \sum \vl_{i} \right),
\eeq
and analogously for the $C$-odd projections $\tilde B_{N;0\,}.$


In the remaining part of this section the main emphasis will be on the color 
structure of the impact factors, contained in Eq.~(\ref{colortrace}).
As the main result, we will find a decomposition into a sum of terms which, as 
it will be demonstrated in the subsequent section, stays invariant under 
evolution in rapidity. We stress that the results which follow are 
valid for an arbitrary overlap function $\;{F}(\vl_1,\vl_2,\vl_3)\;$, i.e. 
they do not rely on a particular model of the baryon wave function, 
provided the baryon has the valence degrees of freedom of three quarks.

\subsection{Two gluons}

We begin with the two-gluon coupling which is $C$-even. All diagrams are 
proportional the color tensor $\delta^{a_1 a_2}$, and it is suggestive to 
group them into three sets: in the first one, the two gluons couple to the 
quark pair $(12)$, and quark $3$ acts as a spectator (Fig.~12). 
\begin{figure}[t]
\centerline{
\begin{tabular}{ll}
\epsfig{file=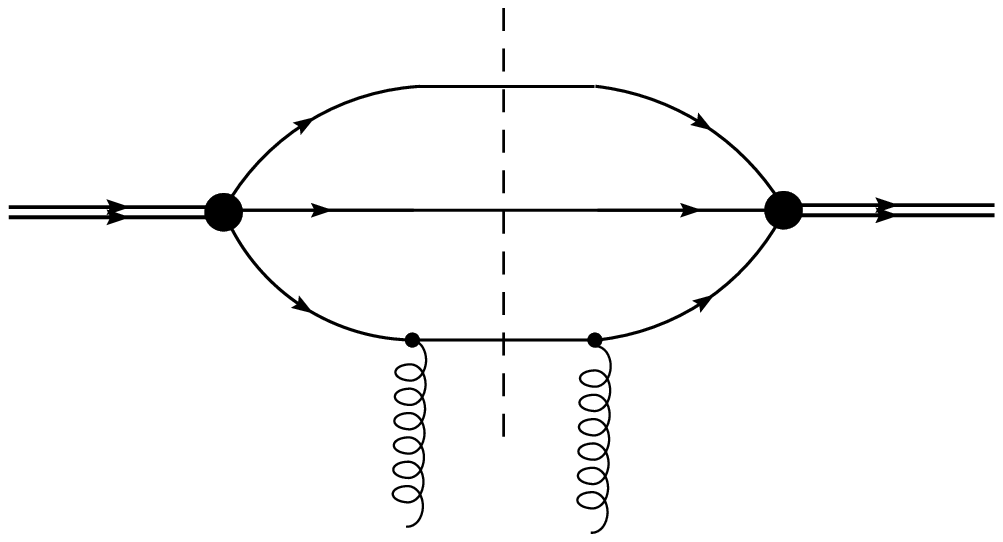,width=3.5cm} &
\epsfig{file=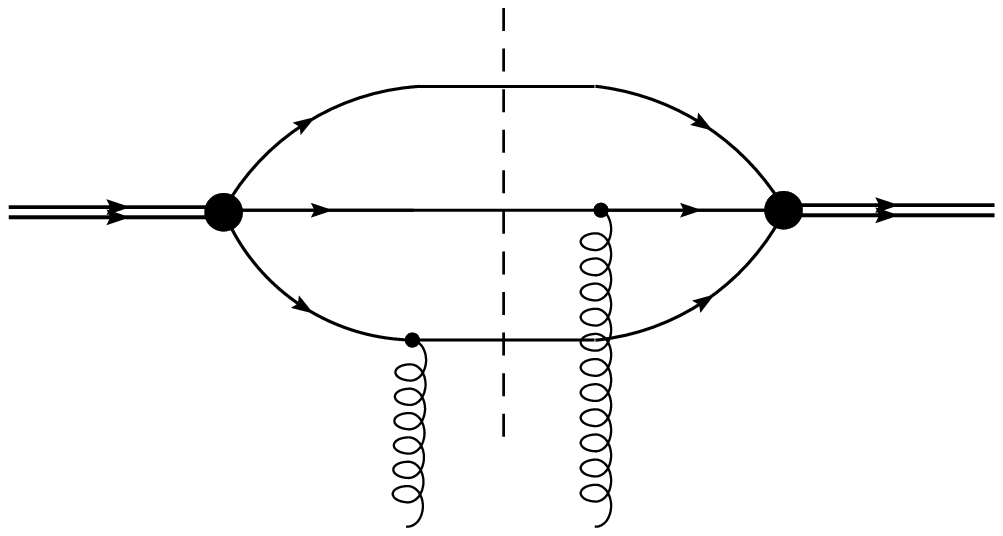,width=3.5cm} \\
{\large\bf a)} & {\large\bf b)} \\
\end{tabular}
}
\caption{\it Diagrams defining $B_{2;0}$.
\label{fig:2g} } 
\end{figure}
In the second one the gluons couple to $(13)$ and quark $2$ acts 
as spectator etc. Inside each set, we have four terms. We thus find:
\beq
B_{2;0} (\vk_1,\vk_2)  =  {\delta\,}^{a_1 a_2}\;
\left[
\Ba_{2;0} ^{\{1,2\}} (\vk_1,\vk_2) +
\Ba_{2;0} ^{\{1,3\}} (\vk_1,\vk_2) +
\Ba_{2;0} ^{\{2,3\}} (\vk_1,\vk_2)\,
\right], 
\label{dipole}
\eeq
with 
\beeq
\Ba_{2;0} ^{\{1,2\}} (\vk_1,\vk_2) & = & 
{-g^2\over 12}\; [ 
F(\vk,0,0) +  F(0,\vk,0)
- F(\vk_1,\vk_2,0) -  F(\vk_2,\vk_1,0)
],
\label{D_2,12} \\
\Ba_{2;0} ^{\{1,3\}} (\vk_1,\vk_2) & = & 
{-g^2\over 12}\; [
F(\vk,0,0)+  F(0,0,\vk)
- F(\vk_1,0,\vk_2) -  F(\vk_2,0,\vk_1)
],\label{D_2,13}
\\
\Ba_{2;0} ^{\{2,3\}} (\vk_2,\vk_3) & = & 
{-g^2\over 12} \;[
F(0,\vk,0) +  F(0,0,\vk)
- F(0,\vk_1,\vk_2) - F(0,\vk_2,\vk_1)
],
\label{D_2,23}
\eeeq
where $\vk_1$, $\vk_2$ denote the gluon momenta and $\vk= \vk_1 + \vk_2$.
On the r.h.s.\ of (\ref{D_2,12})---(\ref{D_2,23}), the momentum arguments 
of the $F$ functions indicate which diagrams they represent: in the  
first (second) term of (\ref{D_2,12}), both gluons couple to quark line 
$1$ ($2$). 
In the third term, the first gluon couples to line $1$, the second to 
line $2$, and so on. The relative signs arise from the color structure.
As a striking result, on the r.h.s.\ of (\ref{D_2,12})---(\ref{D_2,23}), 
in each line the four terms have the same structure as 
the impact factor of the photon. 
In particular, each set satisfies the Ward identities, i.e.\ it 
vanishes as any of its momenta goes to zero.  
Since the pair of scattering quarks $\{i,j\}$ is in a color anti-triplet 
state, one might, at first sight, interpret this set as the elastic scattering of 
an `anti-triplet dipole'. 
However, it is important to stress that these three dipole-like components 
$\Ba_{2;0} ^{\{i,j\}}$, are not independent from each other: 
the diagrams where two gluons couple to the same quark line, say, line~$3$ in Fig.~12a,  
contribute both to the pair $(13)$ and $(23)$. In this sense, one better 
views these quark pairs as `anti-triplets inside the baryon'. 
Also, these configurations where
one quark pair interacts whereas the third quark remains a spectator, 
should not simply be viewed as `diquark states': in transverse coordinate 
space, the spectator quark can be far away from the quark pair 
(see the discussion in Section 7). 
One should also add that the normalization of the dipole-like components 
$\Ba_{2;0} ^{\{i,j\}}$ of the baryon impact factor is exactly $1/2$ of 
the normalization of the genuine color dipole impact factor. At the two-gluon
level, our results coincide with results of Ref.\ \cite{kov_bar}.

If, instead of our model for the baryonic impact factor, we would have used 
a completely symmetric baryon form-factor $F^{(s)}$ 
(which does not discriminate between $u$ and $d$ quarks) 
we would have arrived at a familiar result
\cite{fuk_kwie}:
\beq
B_{2;0} ^{(s)} (\vk_1,\vk_2) \, = \, {-g^2 \over 2} \,\delta^{a_1 a_2} \, \left[ 
F^{(s)}(\vk,0,0) - F^{(s)}(\vk_1,\vk_2,0) \right].
\eeq

\subsection{Three gluons}

\begin{figure}[t]
\centerline{
\begin{tabular}{lll}
\epsfig{file=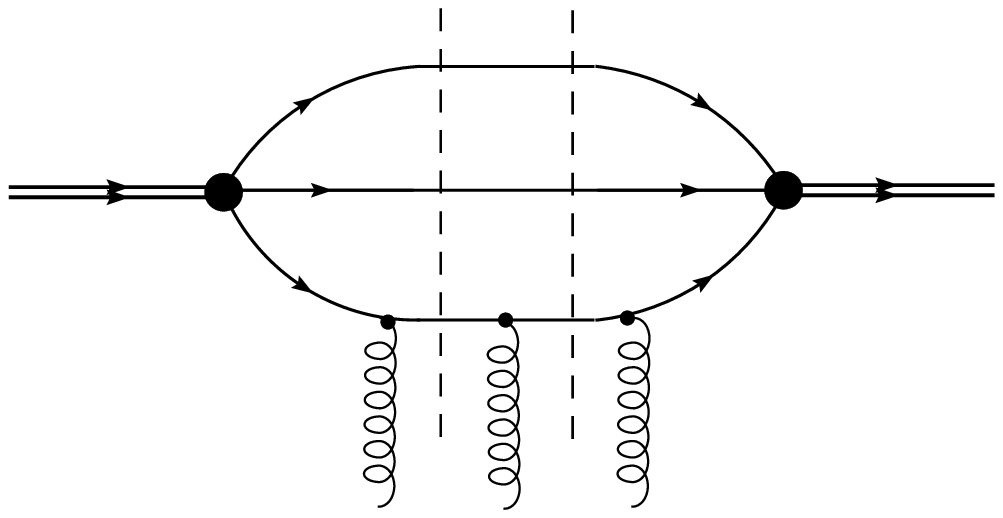,width=3.5cm} &
\epsfig{file=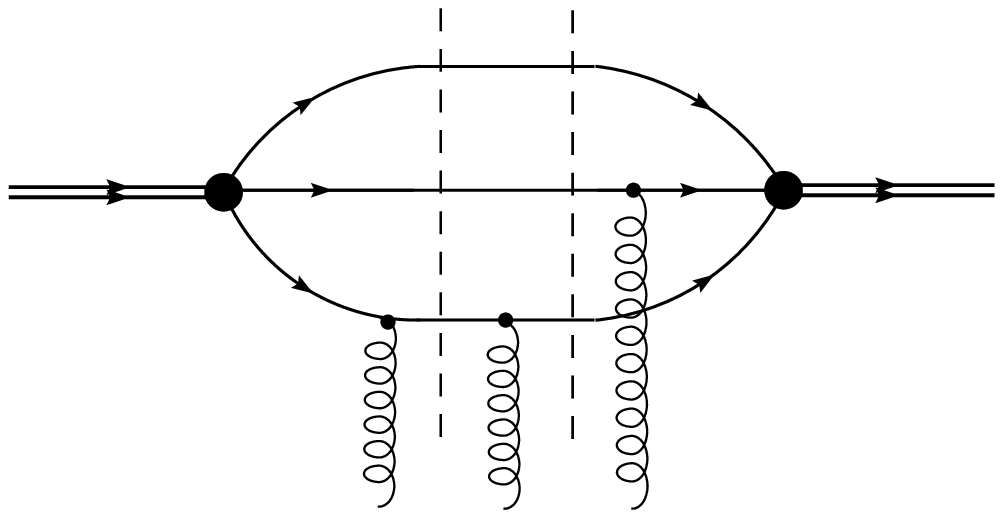,width=3.5cm} &
\epsfig{file=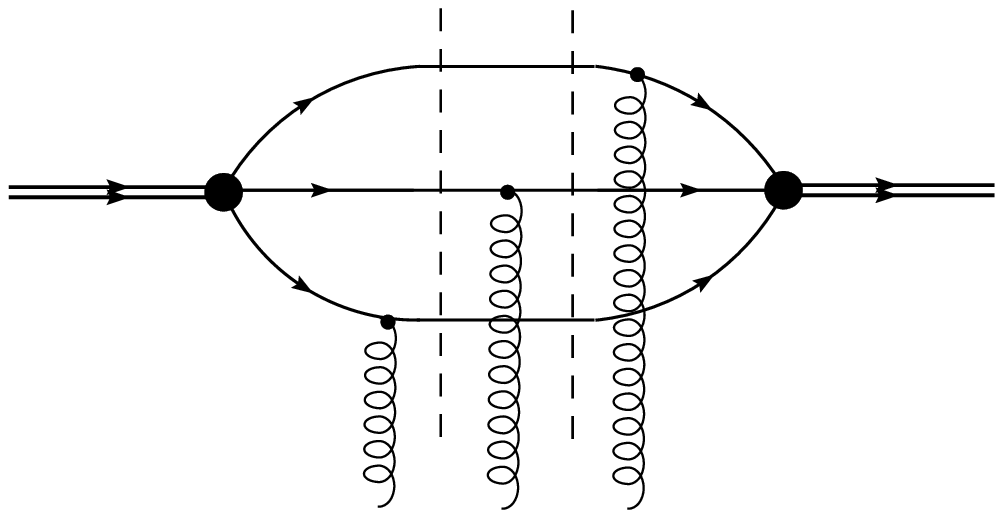,width=3.5cm} \\
{\large\bf a)} & {\large\bf b)} & {\large\bf c)} \\
\end{tabular}
}
\caption{\it Examples of diagrams defining $B_{3;0}$ and $\tilde B_{3;0}$.} 
\end{figure}

In the case of three gluons (Fig.~13) we have to distinguish between even 
and odd $C$ parity: in the color trace Eq.~(\ref{colortrace}) we find both 
color tensors, $f^{a_1 a_2 a_3}$ and $d^{a_1 a_2 a_3}$. 
The first one belongs to even (Pomeron), the second to odd 
(odderon) $C$ parity. 

The $C$-odd baryon impact factor reads~\cite{ewerz_odd,ckms,hiim}
\beq
\tilde B_{3;0}(\vk_1,\vk_2,\vk_3) \, =  \,
{d\,}^{a_1 a_2 a_3} \, E_{3;0}(\vk_1,\vk_2,\vk_3), 
\eeq
where
\beq
E_{3;0} (\vk_1,\vk_2,\vk_3) \; =  \; 
{i g^3\over 24} 
\sum_{\prm} \left[
2F^\prm(\vk_1,\vk_2,\vk_3) - \sum_{i=1} ^3 F^\prm(\vk_i,\vk-\vk_i,0) + F^\prm(\vk,0,0)
\right],
\label{3gluonodd}
\eeq
and $F^{\prm}$ denotes the $F$ functions, with its arguments being  
permuted by the permutation~$\prm$:
\beq
F^\prm(\vl_1,\vl_2,\vl_3) = F\left(
\vl_{\prm(1)},\vl_{\prm(2)},\vl_{\prm(3)}
\right). 
\eeq
In $E_{3;0}$ the $t$-channel three gluon state is Bose symmetric, that is
\beq
E_{3;0}\left(\vk_{\prm(1)},\vk_{\prm(2)},\vk_{\prm(3)}\right) 
\; = \;  E_{3;0}(\vk_1,\vk_2,\vk_3)
\eeq
for any permutation $\prm$, and it obeys the Ward identities:
\beq
E_{3;0}(\vk_1,\vk_2,\vk_3) \; = \; 0\;\;\ \mbox{for any }\;\vk_j \to 0.
\eeq
On the r.h.s.\ of Eq.~(\ref{3gluonodd}), the momentum structure of first 
term indicates that the three gluons couple to three quarks.
The second and third term play the role of subtractions.
This leads to the interpretation that, in this piece of the baryonic impact 
factor, in contrast to the structure found previously for $2$ gluons, 
all three quarks participate in the interaction. Since each of the three 
gluon has negative $C$ parity, this $t$-channel belongs to the 
$C= -$ (odderon) state. 
     
For a completely symmetric model for the baryon form-factor 
expression (\ref{3gluonodd}), again, reduces to a known result~\cite{ckms}:
\beq
\tilde B_{3;0} ^{(s)}(\vk_1,\vk_2,\vk_3) \; =  \;
{i\,g^3 \over 4} \, d^{a_1 a_2 a_3} \,
\left[ 
2F^{(s)}(\vk_1,\vk_2,\vk_3) - \sum_{i=1} ^3 F^{(s)}(\vk_i,\vk-\vk_i,0) + F^{(s)}(\vk,0,0)
\right].
\eeq

Next we turn to the terms proportional to $f^{a_1 a_2 a_3}$ which turn
out to belong to even $C$. They can be grouped in the same `dipole-like'
form as in (\ref{dipole}):   
\beq
B_{3;0}(\vk_1,\vk_2,\vk_3) \; = \; \Ba_{3;0} ^{\{1,2\}} (\vk_1,\vk_2,\vk_3) 
                 \,+\, \Ba_{3;0} ^{\{1,3\}} (\vk_1,\vk_2,\vk_3) 
                 \,+\, \Ba_{3;0} ^{\{2,3\}} (\vk_1,\vk_2,\vk_3), 
\label{3gluon}
\eeq
where the dipole-like components have the structure known from the photon case,
\beq
\Ba_{3;0} ^{\{i,j\}} (\vk_1,\vk_2,\vk_3)  \; = \; {1\over 2}\,g\,f^{a_1 a_2 a_3} \; 
\left[ 
\Ba_{2;0} ^{\{i,j\}} (\vk_1+\vk_2,\vk_3) - 
\Ba_{2;0} ^{\{i,j\}} (\vk_1+\vk_3,\vk_2) + 
\Ba_{2;0} ^{\{i,j\}} (\vk_2+\vk_3,\vk_1) 
\right].
\label{D30}
\eeq
As in the photon case, the argument structure indicates the beginning of the reggeization
of the gluons: for example, in the first term, the first two gluons 
with momenta $\vk_1$ and $\vk_2$ `collapse' into a single reggeized gluon 
with momentum $\vk_1+\vk_2$. The $t$-channel system thus consists of 
two reggeized gluons only and hence belongs to $C$-even. 
In the next section we will show that this structure is preserved in the rapidity
evolution.   

In the following it will be convenient to use a shorthand notation 
by writing, instead of   
$\Ba_{2;0} ^{\{i,j\}} (\vk_1+\vk_2,\vk_3)$, simply 
$\Ba_{2;0} ^{\{i,j\}} (12,3)$ etc.

\subsection{Four gluons}

In the case of four gluons (Fig.~14) the color trace (\ref{colortrace}) 
contains
$ff$, $dd$, $fd$, and $\delta \delta$ color tensor structures. 
Beginning with the $fd$ pieces, we find that they can be expressed 
in terms of the $E$-function 
(\ref{3gluonodd}) which we have obtained for the odderon channel:
\beeq
\tilde B_{4;0}(1,2,3,4) & = &  \\
& & \hspace{-5mm} {g\over 2}\,
\left(
\,
f^{a_1 a_2 b} \, d^{b a_3 a_4} \, E_{3;0}(12,3,4) \,+\,  
f^{a_1 a_3 b} \, d^{b a_2 a_4} \, E_{3;0}(13,2,4) \,+\,
f^{a_1 a_4 b} \, d^{b a_2 a_3} \, E_{3;0}(14,2,3) \nonumber 
\right.\\
& + & 
\left.
\;f^{a_2 a_3 b} \, d^{b a_1 a_4} \, E_{3;0}(23,1,4) \,+\,  
f^{a_2 a_4 b} \, d^{b a_1 a_3} \, E_{3;0}(24,1,3) \,+\,
f^{a_3 a_4 b} \, d^{b a_1 a_2} \, E_{3;0}(34,1,2)\,
\right). 
\nonumber
\eeeq
We then interpret this contribution as the odderon configuration with 
one reggeizing gluon. It agrees with the result first found by C.~Ewerz 
~\cite{ewerz_odd}.

\begin{figure}[h]
\centerline{
\begin{tabular}{llll}
\epsfig{file=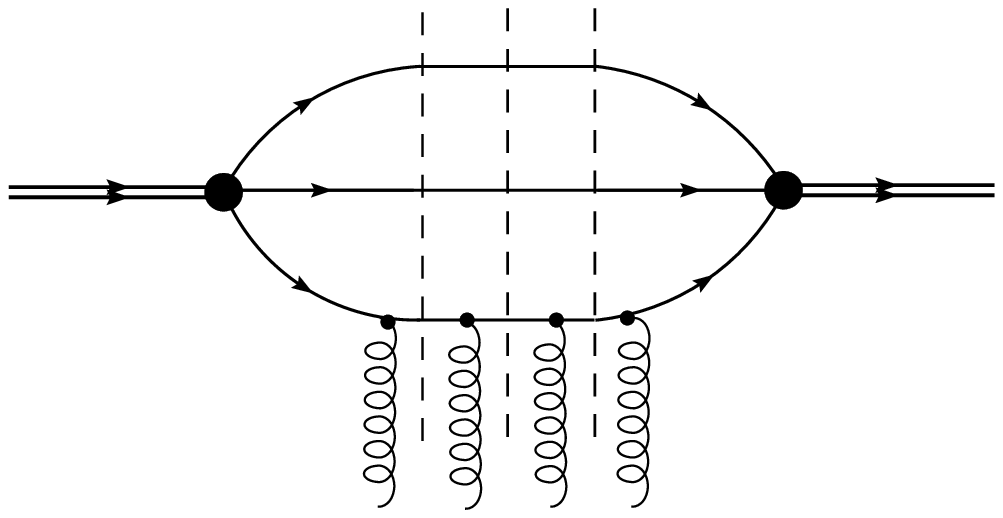,width=3.5cm} &
\epsfig{file=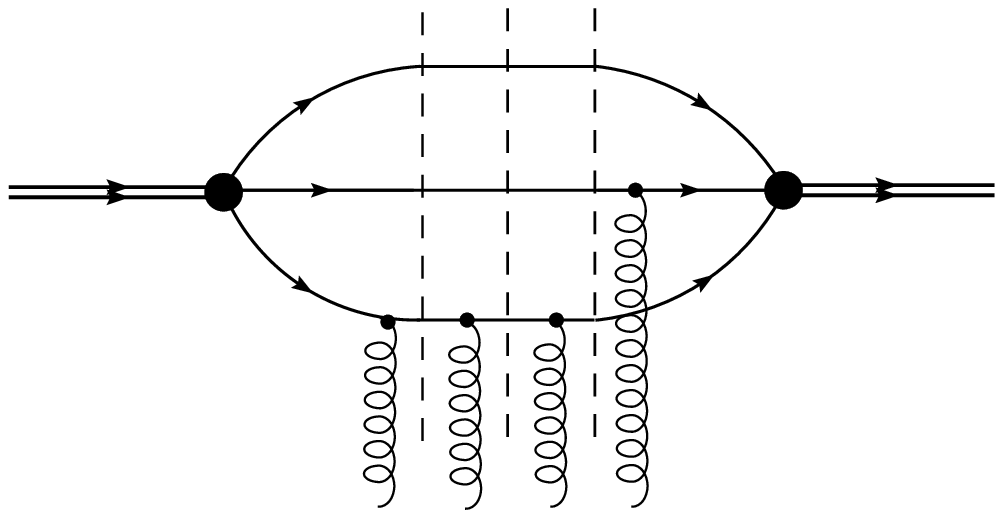,width=3.5cm} &
\epsfig{file=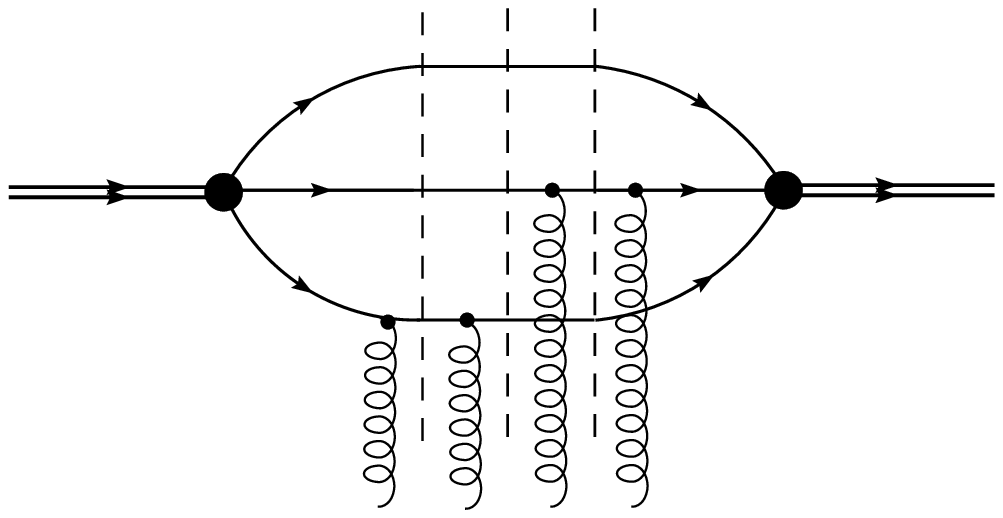,width=3.5cm} &
\epsfig{file=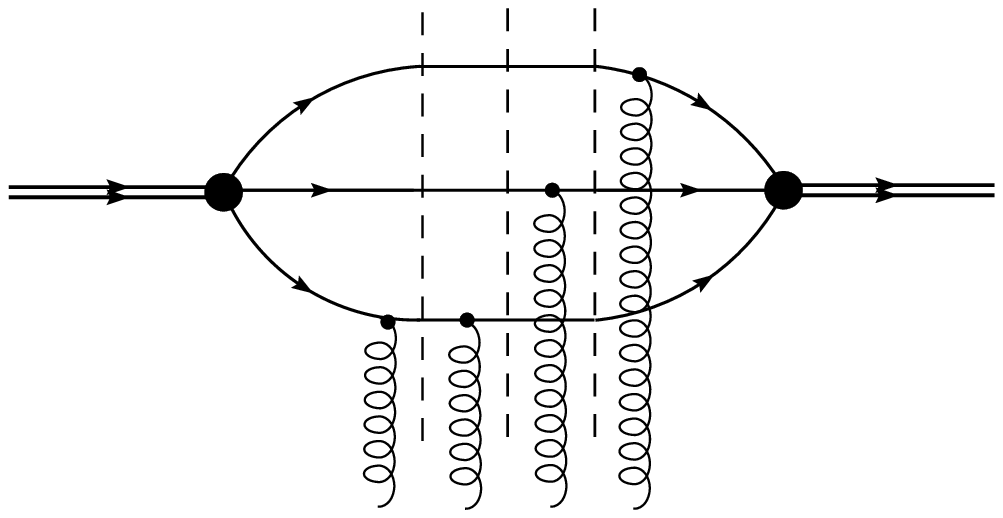,width=3.5cm} \\
{\large\bf a)} & {\large\bf b)} & {\large\bf c)} & {\large\bf d)}\\
\end{tabular}
}
\caption{\it Examples of diagrams defining $B_{4;0}$ and $\tilde B_{4;0}$.} 
\end{figure}

Next the $ff$, $dd$, and $\delta \delta$ terms. 
We find, in addition to a set of pieces which have the same structure 
as in the photon case, a new structure, $Q_{4;0}$. In detail:
\beq
B_{4;0}(1,2,3,4) \;=\; 
D_{4;0} ^{\{1,2\}} ( 1,2,3,4) \,+\,
D_{4;0} ^{\{1,3\}} ( 1,2,3,4) \,+\,
D_{4;0} ^{\{2,3\}} ( 1,2,3,4) \,+\,
Q_{4;0} (1,2,3,4). 
\label{B40}
\eeq
Here the first three terms are dipole-like, and they follow the 
reggeization pattern found for the photon scattering:
\beeq 
D_{4;0} ^{\{i,j\}} (1,2,3,4) & = & 
\; -g^2 \left\{
\, {d\,}^{a_1 a_2 a_3 a_4} \; 
\left[\, \Ba_{2;0}^{\{i,j\}}(123,4) +  \Ba_{2;0}^{\{i,j\}} (234,1) 
- \Ba_{2;0}^{\{i,j\}}(14,23) \,\right]  \right. \\
 & & 
\left.
+ \; {d\,}^{a_1 a_2 a_4 a_3} \; 
\left[\, \Ba_{2;0}^{\{i,j\}}(124,3) +  \Ba_{2;0}^{\{i,j\}}(134,2) 
  - \Ba_{2;0}^{\{i,j\}}(12,34) - \Ba_{2;0}^{\{i,j\}}(13,24)\, \right]
\right\}, \nonumber
\eeeq
with the color tensor
\beq
{d\,}^{a_1 a_2 a_3 a_4} \, = \, 
{{\delta\,}^{a_1 a_2} {\delta\,}^{a_3 a_4} \over 2N_c} \, + \,
{{d\,}^{a_1 a_2 b} \, {d\,}^{b a_3 a_4} \over 4}  \, - \,
{{f\,}^{a_1 a_2 b} \, {f\,}^{b a_3 a_4} \over 4}. 
\eeq
In the next section we will study the rapidity evolution of these terms, and 
we will confirm that they follow the photon impact factor to all orders. 

The new structure which has no analogue in the case of the 
photon looks as follows:
\beeq
\label{eq:q40}
Q_{4;0}(1,2,3,4) & = & 
{-ig\over 2}\; \left[\, 
{d\,}^{a_1 a_2 b}\, {d\,}^{b a_3 a_4} 
\, - \, \frac{1}{3}\, {\delta\,}^{a_1 a_2} \, {\delta\,}^{a_3 a_4} 
\right]
\left[\, E_{3;0} (12,3,4) \,+\,  E_{3;0} (34,1,2)\, \right] \;
+ \nonumber \\
& & 
{-ig\over 2}\; \left[\, 
{d\,}^{a_1 a_3 b}\, {d\,}^{b a_2 a_4} \, - \,
\frac{1}{3}\, {\delta\,}^{a_1 a_3} \, {\delta\,}^{a_2 a_4} 
\right]
\left[\, E_{3;0} (13,2,4) \,+\,  E_{3;0} (24,1,3)\, \right]\;
+ \nonumber \\
& & 
{-ig\over 2}\; \left[\, 
{d\,}^{a_1 a_4 b}\, {d\,}^{b a_2 a_3} \, - \,
\frac{1}{3}{\delta\,}^{a_1 a_4} \, {\delta\,}^{a_2 a_3} 
\right]
\left[\, E_{3;0} (14,2,3) \,+\,  E_{3;0} (23,1,4)\, \right].
\eeeq
The function $E$ is the same as in the odderon case, and, in particular,  
all three quarks participate in the interaction.   
The $t$-channel state which couples to $Q_{4;0}$ is Bose symmetric
\beq
Q_{4;0}(\prm(1),\prm(2),\prm(3),\prm(4)) \; = \;  Q_{4;0}(1,2,3,4)
\eeq
for any permutation $\prm$, and it is gauge invariant:
\beq
Q_{4;0}(\vk_1,\vk_2,\vk_3,\vk_4) \; = \; 0\;\;\ \mbox{for any }\;
\vk_{j} \to 0. \eeq
This property may be proven using the identity for color tensors
valid for $\,N_c = 3\,$:
\beq
\label{eq:deltad}
d^{\,a_1 a_2 b}\, d^{\,b a_3 a_4} \,+\, 
d^{\,a_1 a_3 b}\, d^{\,b a_2 a_4} \,+\,
d^{\,a_1 a_4 b}\, d^{\,b a_2 a_3} \;\; = \;\;
{1\over 3}\left(\,
\delta^{\,a_1 a_2} \, \delta^{\,a_3 a_4} \, + \,
\delta^{\,a_1 a_3} \, \delta^{\,a_2 a_4} \, + \,
\delta^{\,a_1 a_4} \, \delta^{\,a_2 a_3}\,\right). 
\eeq
The analysis in the following section will show that this novel piece of the  
baryon impact factor couples a three-gluon $t$-channel configuration 
in which one of the reggeized gluons is an even-signature $d$-Reggeon.
The overall $C$ parity therefore is positive.

\section{Integral evolution equations}

In this section we study higher order corrections in the (generalized) 
leading logarithmic ($\log s$) approximation. The all-order sum of these terms will be 
represented by integral equations~\cite{bw,be}, written for Mellin moments
of the multiple discontinuities with respect to the energy $s$. In our notation 
the dependence of the amplitudes $B_N$ and $\tilde B_N$ (and also $D_N$, $E_N$ and
$Q_N$) on the Mellin variable $\omega$ is implicit.

Let us begin with the $C$-odd configurations. In the case of three gluons,
the impact factor $E_{3;0}$ is simply replaced by the Green's function function $E_{3}$, which satisfies 
the BKP equation for three odd signature Reggeons, 
with the initial condition given by $E_{3;0}$:
\beq 
\left(\omega - \sum_i \beta(\vk_i) \right) \, E_3 \; = \; E_{3;0} \; + \;
\sum_{(r,s)} \; K_{2\to 2}(r,s) \, \otimes \, E_3,
\label{E3}
\eeq  
where $K_{2\to 2}$ is the real emission part of the BFKL kernel, and
the odderon state with the full color structure reads
\beq
\tilde B_3(1,2,3) \; = \; d^{a_1 a_2 a_3} E_3(1,2,3).
\eeq
The four gluon case has been studied in ~\cite{ewerz_odd},
and we simply quote the solution:
\beeq
\tilde B_{4}(1,2,3,4) & = &  \\
& & \hspace{-6mm}{g\over 2} \, \left[ \,
f^{a_1 a_2 b} \, d^{b a_3 a_4} \, E_{3}(12,3,4) \,+\,  
f^{a_1 a_3 b} \, d^{b a_2 a_4} \, E_{3}(13,2,4) \,+\,
f^{a_1 a_4 b} \, d^{b a_2 a_3} \, E_{3}(14,2,3) 
\right.
\nonumber \\
& + & 
\left.
f^{a_2 a_3 b} \, d^{b a_1 a_4} \, E_{3}(23,1,4) \,+\,  
f^{a_2 a_4 b} \, d^{b a_1 a_3} \, E_{3}(24,1,3) \,+\,
f^{a_3 a_4 b} \, d^{b a_1 a_2} \, E_{3}(34,1,2)\,
\right]. \nonumber
\eeeq
where the function $E_3$ has been defined before in (\ref{E3}). 
Clearly, the solution is saturated by a reggeizing contribution: in each term, 
one of the three $f$-Reggeons splits into two elementary gluons.

We now turn to the $C$-even contributions. The integral equations for the multiple discontinuities 
read (up to four gluons):  
\beeq
\left(\omega - \sum_i \beta(\vk_i) \right) 
\raisebox{-8mm}{\epsfig{file=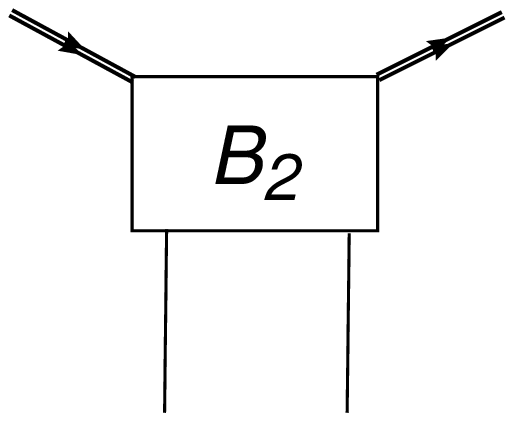,width=22mm}} & = &
\raisebox{-8mm}{\epsfig{file=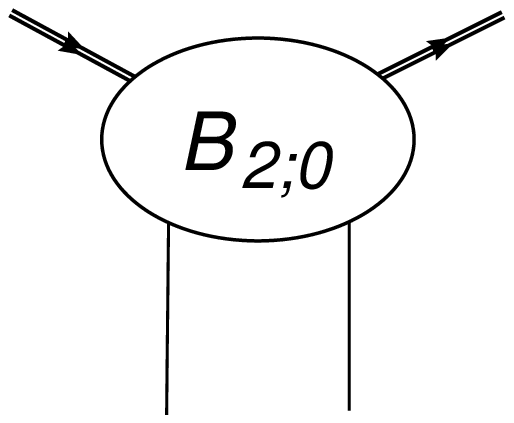,width=22mm}} \;+\; 
\raisebox{-8mm}{\epsfig{file=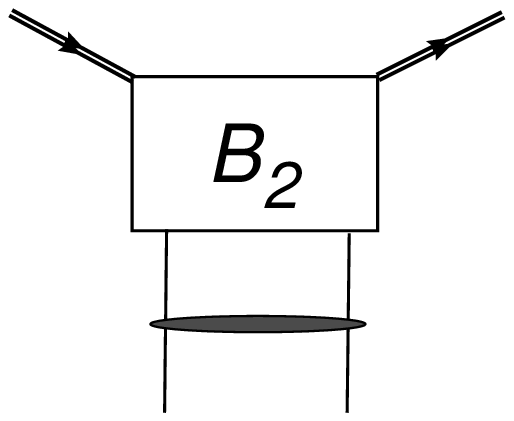,width=22mm}} 
\label{B2}
\\
\left(\omega - \sum_i \beta(\vk_i) \right) 
\raisebox{-8mm}{\epsfig{file=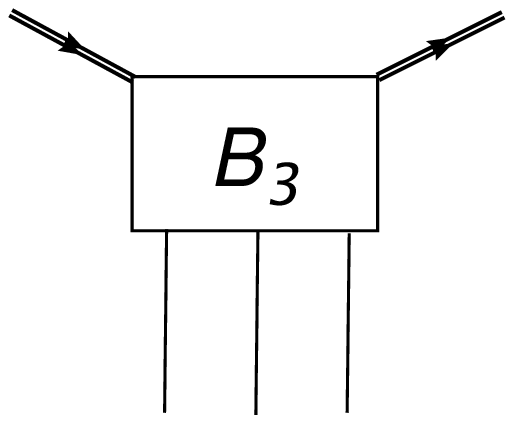,width=22mm}} & = &
\raisebox{-8mm}{\epsfig{file=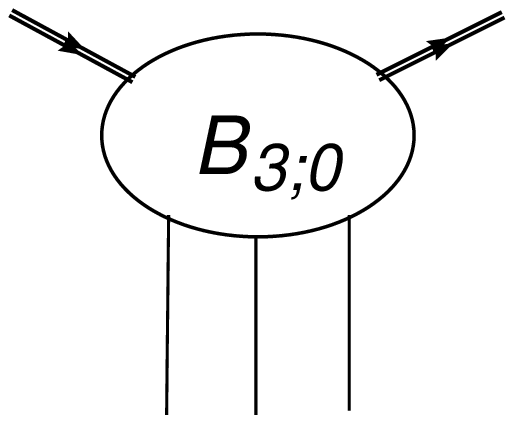,width=22mm}} \;+\; 
\sum \; \raisebox{-8mm}{\epsfig{file=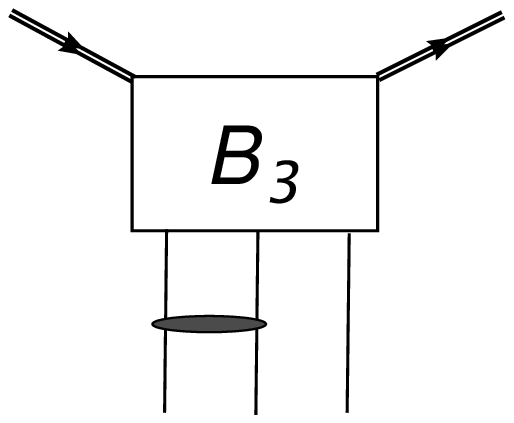,width=22mm}} \;+\;
\raisebox{-8mm}{\epsfig{file=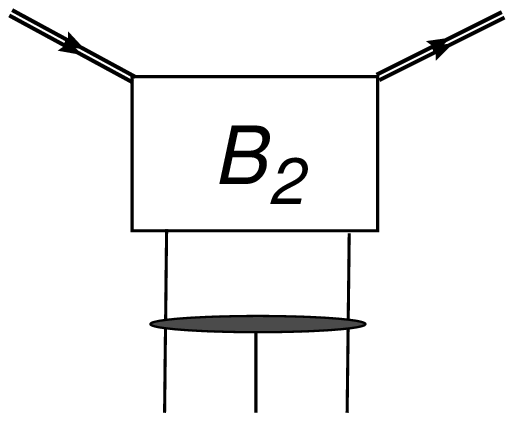,width=22mm}} \label{B3}\\
\left(\omega - \sum_i \beta(\vk_i) \right) 
\raisebox{-8mm}{\epsfig{file=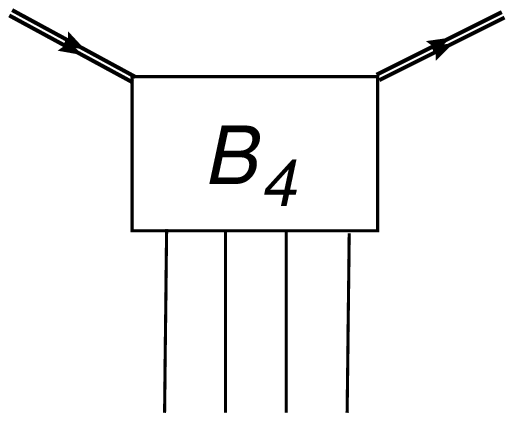,width=22mm}} & = &
\raisebox{-8mm}{\epsfig{file=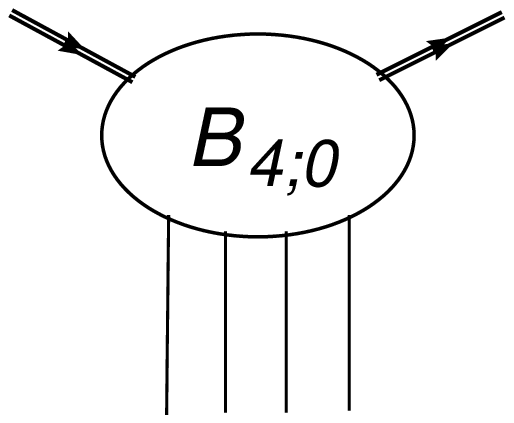,width=22mm}} \;+\; 
\sum\; \raisebox{-8mm}{\epsfig{file=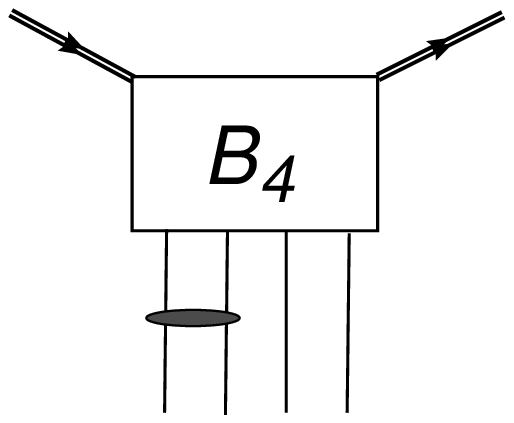,width=22mm}} \;+\;
\sum\; \raisebox{-8mm}{\epsfig{file=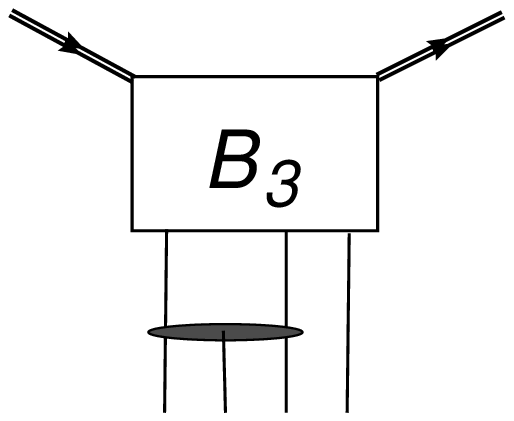,width=22mm}}  \;  \nonumber \\
& + &  \raisebox{-8mm}{\epsfig{file=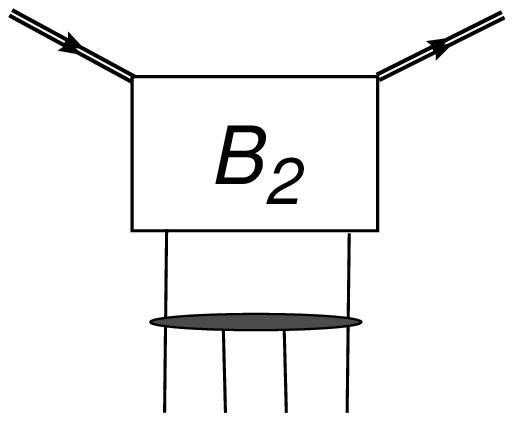,width=22mm}}. \label{B4}
\eeeq
The integral kernels driving $2 \to 2,3,4,...$ Reggeon transitions, 
that appear in the above equations include the color structure, and 
they are defined in Ref.~\cite{bw,be}. The gluon Regge trajectory 
$\beta(\vk)$ will be specified below.
The case of two gluons is the simplest one: $B_2$ satisfies the BFKL equation.
According to the structure of the inhomogeneous term in Eq.\ (\ref{dipole}), 
$B_2$ can be written as the sum of three terms $D_2^{\{i,j\}}$,
\beq
B_2(1,2) \; = \; \delta^{a_1 a_2}\, \left[\, 
D_2^{\{1,2\}}(1,2) \,+\, D_2^{\{1,3\}}(1,2) \,+\,  D_2^{\{2,3\}}(1,2)\, \right],
\eeq
with    
\beq 
\left(\omega - \sum_{i=1} ^2 \beta(\vk_i) \right) \, D_2 ^{\{i,j\}} \; = \; 
D^{\{i,j\}}_{2;0} \; + \;  K_{2\to 2} \, \otimes \, D_2 ^{\{i,j\}}.
\eeq

In the case of three gluons, $B_3$ is given by the sum of three dipole-like 
components  (cf.\ (\ref{3gluon})):  
\beq
B_3(1,2,3) \; = \;  
\Ba_{3} ^{\{1,2\}}(1,2,3) \,+\, 
\Ba_{3} ^{\{1,3\}}(1,2,3) \,+\, 
\Ba_{3} ^{\{2,3\}}(1,2,3), 
\eeq 
where each term consists of three reggeizing pieces: 
\beq
\Ba_{3} ^{\{i,j\}} (1,2,3)  \; = \;  {1\over 2}\,g\,f^{a_1 a_2 a_3} \; 
\left[ 
\Ba_{2} ^{\{i,j\}} (12,3) - 
\Ba_{2} ^{\{i,j\}} (13,2) + 
\Ba_{2} ^{\{i,j\}} (23,1) \right].
\eeq
This structure coincides with the photon case.

The case of $B_4$ is more complex. Following our result for the baryon impact 
factor in Eq.\ (\ref{B40}) we decompose $B_4$ in the following way:
\beq
B_{4}(1,2,3,4) \;=\; 
D_{4} ^{\{1,2\}} ( 1,2,3,4) \,+\,
D_{4} ^{\{1,3\}} ( 1,2,3,4) \,+\,
D_{4} ^{\{2,3\}} ( 1,2,3,4) \,+\,
Q_{4} (1,2,3,4). 
\label{fullB4}
\eeq
For the dipole-like pieces $D_4^{\{i,j\}\,}$ we make use of the 
`reduction procedure' developed for the photon case. 
Namely we decompose each 
$D_4^{\{i,j\}}$ into a reggeizing and an irreducible contributions
\beq 
D_{4}^{\{i,j\}\,} (1,2,3,4) \;\; = \;\; 
D_{4} ^{\{i,j\}\, ;R} (1,2,3,4) \; + \;
D_{4} ^{\{i,j\}\, ;I} (1,2,3,4), 
\eeq
with the reggeizing contribution given by   
\beeq 
D_{4} ^{\{i,j\}\, ;R} (1,2,3,4) & = & 
-g^2\,\left\{ {d\,}^{a_1 a_2 a_3 a_4} \; 
\left[\, D_{2}^{\{i,j\}}(123,4) +  D_{2}^{\{i,j\}} (234,1) 
- D_{2}^{\{i,j\}}(14,23) \,\right]\right.   \\
 & & \hspace{-5mm} \left. + \, {d\,}^{a_1 a_2 a_4 a_3} \; 
\left[\, D_{2}^{\{i,j\}}(124,3) +  D_{2}^{\{i,j\}}(134,2) 
  - D_{2}^{\{i,j\}}(12,34) - D_{2}^{\{i,j\}}(13,24)\, \right] \right\}. 
\nonumber
\eeeq
The reggeizing contributions are simple BFKL ladders with one reggeizing gluon splitting into three 
gluons or both reggeized gluons each splitting into two gluons.
The irreducible contribution, containing the $2 \to 4$ Reggeon transition 
vertex, is illustrated in Fig.~15.

\begin{figure}[h]
\leavevmode
\begin{center}
\epsfig{file=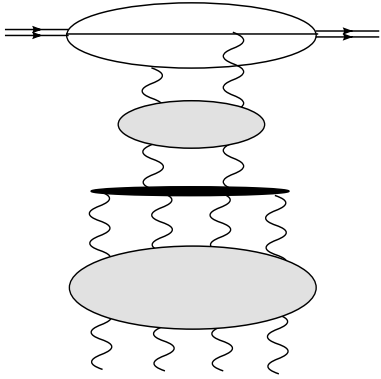,width=5cm,height=5cm}\\
\end{center}
\caption{\it The component $D_{4} ^{\{i,j\}\, ;I} (1,2,3,4)$. }
\end{figure}
These results provide further evidence that the `dipole-like' 
pieces of the baryonic impact factor really behave in exactly the same way as 
the color dipole photon impact factor. In particular, if we would apply the 
large $N_c$ limit 
to the gluon evolution below the impact factor (which, of course, 
would be inconsistent with our finite-$N_c$ baryon), the four gluon system 
below the $2 \to 4$ transition vertex would split into two non-interacting 
BFKL ladders, and we would arrive at the first iteration of the BK equation.      

After subtracting, from $B_4(1,2,3,4)$ in (\ref{fullB4}), these dipole-like 
contributions of the baryon we are left with $Q_4$. As $Q_4$ appears at level 
of four gluons, its evolution equation has simply the BKP form: 
\beq
\left(\omega - \sum_i \beta(\vk_i) \right) 
\raisebox{-8mm}{\epsfig{file=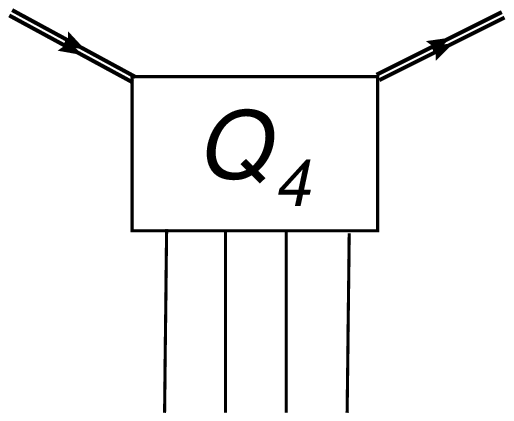,width=22mm}}  = 
\raisebox{-8mm}{\epsfig{file=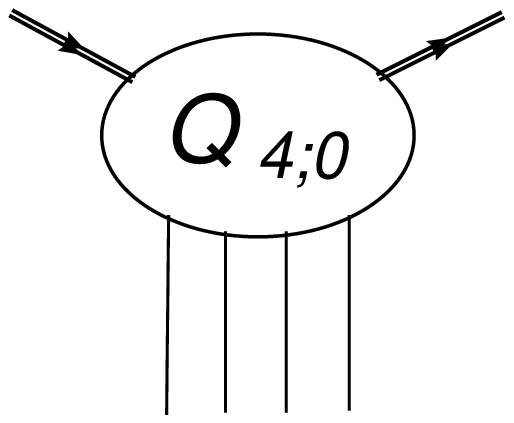,width=22mm}} \;+\; 
\sum\; \raisebox{-8mm}{\epsfig{file=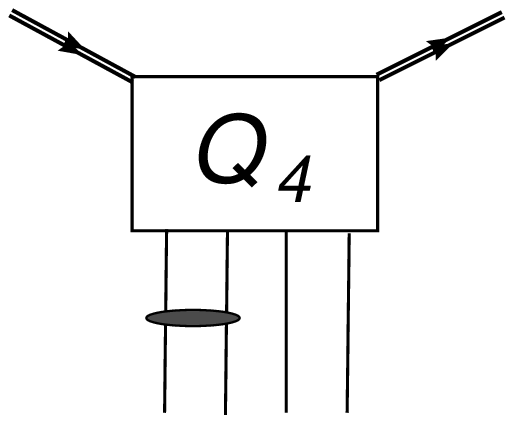,width=22mm}}. 
\eeq
Making use of the experience with $D_4$, we decompose the amplitude 
$Q_4$ into a reggeizing piece  $Q_{4} ^{R}$  and an irreducible contribution
$Q^I _{4}$: 
\beq 
Q_4 (1,2,3,4) \;\; = \;\; Q_{4} ^{R} (1,2,3,4) \;+\;
Q_{4} ^{I} (1,2,3,4). 
\eeq
Going through steps similar to the ones outlined in 
~\cite{bw,be} we find that the reggeizing pieces $Q_{4} ^{R}$ preserve the 
structure of  $Q_{4;0}$:
\beeq
Q_{4} ^R (1,2,3,4) & = & 
{-ig\over 2}\; \left[\, 
{d\,}^{a_1 a_2 b}\, {d\,}^{b a_3 a_4} 
\, - \, \frac{1}{3}\, {\delta\,}^{a_1 a_2} \, {\delta\,}^{a_3 a_4} 
\right]
\left[\, E_{3} (12,3,4) \,+\,  E_{3} (34,1,2)\, \right] \;
+ \nonumber \\
& & 
{-ig\over 2}\; \left[\, 
{d\,}^{a_1 a_3 b}\, {d\,}^{b a_2 a_4} \, - \,
\frac{1}{3}\, {\delta\,}^{a_1 a_3} \, {\delta\,}^{a_2 a_4} 
\right]
\left[\, E_{3} (13,2,4) \,+\,  E_{3} (24,1,3)\, \right]\;
+ \nonumber \\
& & 
{-ig \over 2}\; \left[\, 
{d\,}^{a_1 a_4 b}\, {d\,}^{b a_2 a_3} \, - \,
\frac{1}{3}{\delta\,}^{a_1 a_4} \, {\delta\,}^{a_2 a_3} 
\right]
\left[\, E_{3} (14,2,3) \,+\,  E_{3} (23,1,4)\, \right].
\eeeq
As seen from the color and momentum structure, the three gluon state 
coupling to $Q_{4;0}$ consists of three reggeized 
gluons, one of which is in a $d$ state and decays into two elementary 
gluons (the pieces proportional to color tensors $\delta \delta$ play the 
role of subtractions; in particular, they are needed in order to satisfy 
the Ward identities). This state, consisting of two odd signature 
$f$-Reggeon and one even signature $d$-Reggeon, belongs to even $C$, i.e.\ to 
the Pomeron channel.  

The remaining piece, $Q_4^I$, contains a new transition vertex. 
We illustrate this contribution in Fig.~16. This vertex describes the   
\begin{figure}
\begin{center}
\epsfig{file=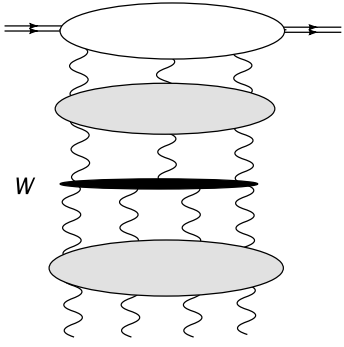, width=4.5cm,height=4.5cm}\\
\end{center}
\caption{\it The new $3 \to 4$ transition vertex $W$.} 
\end{figure}
transition from the three Reggeon state consisting of two $f$ and one $d$ 
Reggeon  to four $f$ Reggeons. In more detail, the vertex may be completely
decomposed into non-connected pieces of two types: 
(i) the incoming $d$ Reggeon together with one of the $f$ Reggeons makes 
a transition into three $f$ Reggeons whereas the remaining $f$ Reggeon acts 
as a ($t$-channel) spectator, and (ii): two $f$~Reggeons interact via the 
BFKL kernel and the $d$~Reggeon splits into two $f$~Reggeons.
The explicit functional form of the vertex $W$, acting on the three Reggeon 
state $\phi_3$ is the following:
\beq
(W\,\phi_3)(1,2,3,4) \; = \; {-g^2\over 2} \left[ \;
{\delta\,}^{a_1 a_2}\, {\delta\,}^{a_3 a_4}\,
\overset{1\;2\;3\;4}{({\cal W}\,\phi_3)} \; + \; 
{\delta\,}^{a_1 a_3}\, {\delta\,}^{a_2 a_4}\,
\overset{1\;3\;2\;4}{({\cal W}\,\phi_3)} \; + \; 
{\delta\,}^{a_1 a_4}\, {\delta\,}^{a_2 a_3}\,
\overset{1\;4\;2\;3}{({\cal W}\,\phi_3)} \;\right],
\label{vertexW}
\eeq
where
\beeq
\overset{1\;2\;3\;4}{({\cal W}\,\phi_3)} & =  & \left[\; 
\overset{123}{\cal G}\, \overset{\cdot \cdot 4}{\phi_3}  \; + \;
\overset{213}{\cal G}\, \overset{\cdot \cdot 4}{\phi_3}  \; + \;
\overset{132}{\cal G}\, \overset{\cdot \cdot 4}{\phi_3}  \; - \;
\overset{(12)\circ 3}{\cal G}\, \overset{\cdot \cdot 4}{\phi_3} \;  
+ \; \frac{1}{2}\overset{1\circ 2}{\cal G}\,\overset{\cdot \cdot (34)}{\phi_3} \; \right] 
\; \nonumber \\ & & +\; [\,3 \leftrightarrow 4\,] \; + \;  
[\,1 \leftrightarrow 3\,,\,2 \leftrightarrow 4\,] \; + \;  
[\,1 \leftrightarrow 4\,,\,2 \leftrightarrow 3\,].
\label{arguments}
\eeeq
Let us stress that this vertex acts on a completely symmetric function 
$\phi_3$ with three arguments, which inherits its structure from $E_3$.  
This vertex is closely related to a $3\to 4$ vertex found in Ref.~\cite{salv} in 
an analysis of jet production amplitudes at small~$x$.
The symbol $\overset{123}{\cal G}$ denotes the integral operator ${G}(1,2,3)$, 
introduced first in~\cite{bw} and further investigated in~\cite{bra_vac}. 
It acts on a two gluon amplitude, $\phi_2$, and describes 
a transition to three gluons. It consists of two pieces:
\beeq
{G}(1,2,3) \; = \; {G}_1(1,2,3) \,+\, {G}_2(1,2,3),
\eeeq
where the first one contains $s$-channel gluons (`connected part'), the 
second one reggeizing pieces (`disconnected part'):
\beq
({G}_1 \phi_2)(\vk_1,\vk_2,\vk_3) \; = \; \nonumber
\eeq
\beeq
\int \frac{d^2 l}{(2\pi)^3} 
\left( 
\frac{(\vk_2+\vk_3)^2\,\vl^2}{(\vl-\vk_1)^2}
\, + \,
\frac{(\vk_1+\vk_2)^2\, (\vk-\vl)^2}
{(\vk-\vl-\vk_3)^2}\,-\,
\frac{\vk_2^2\, (\vk-\vl)^2\, \vl^2}{(\vl-\vk_1)^2\,(\vk-\vl-\vk_3)^2}
\, -\, \vk^2
\right) \phi_2(\vl,\vk-\vl), 
\eeeq
and 
\beq
N_c \, g^2\,({G}_2 \phi_2)(\vk_1,\vk_2,\vk_3) \; = \; \nonumber  
\eeq
\beeq
\int \frac{d^2l}{(2\pi)^3}\, \vl^2 \, (\vk-\vl)^2 \,
\left\{ \,
[\,\beta(\vk_2+\vk_3)-\beta(\vk_2)\,]\; 
(2\pi)^3 \, \delta^{(2)}(\vl-\vk_1)
\right.
\nonumber \\
\left.
+ \;[\, \beta(\vk_1+\vk_2)\, -\, \beta(\vk_2)\, ]\; 
(2\pi)^3 \, \delta^{(2)}(\vl-\vk_3) \,
\right\}\, 
\phi_2(\vl,\vk-\vl),
\eeeq
with the gluon trajectory function
\beq
\beta(\vk_i)\; = \; -N_c g^2\int\frac{d^2\vl}{(2\pi)^3}
\frac{\vk_i^2}{\vl^2+(\vk_i-\vl)^2}\frac{1}{(\vk_i-\vl)^2},
\eeq
and $\vk = \vk_1 + \vk_2 + \vk_3$.
In (\ref{arguments}) we have used a short-hand notation for the argument 
structure introduced in Ref.~\cite{salv}: 
in the first term, $\overset{123}{\cal G}\, \overset{\cdot \cdot 4}{\phi_3}$, 
$\phi_3$ is the three gluon amplitude above the vertex $W$ where the rightmost 
Reggeon (momentum $\vk_4$) is a spectator, and the $G$ operator acts on the 
two left Reggeons, turning them into the three gluons with momenta 
$\vk_1$, $\vk_2$, and $\vk_3$. In the fourth term,      
$\overset{(12)\circ 3}{\cal G}\, \overset{\cdot \cdot 4}{\phi_3}$,
Reggeon~$4$ is, again, a spectator, and the $G$ operator  
(with zero momentum in the second outgoing gluon) equals the BFKL 
kernel acting on the two leftmost gluons inside $\phi_3$: 
after this BFKL interaction the leftmost gluon splits into two gluons 
with momenta $\vk_1$ and $\vk_2$, and the other one carries momentum $\vk_3$. 
Finally, in the last term,
$\overset{1\circ 2}{\cal G}\,\overset{\cdot \cdot (34)}{\phi_3}$, 
the rightmost spectator now splits into two gluons with momenta 
$\vk_3$ and $\vk_4$, and the $G$ operator, like in the previous term,
equals the BFKL operator with outgoing momenta $\vk_1$ and $\vk_2$.         

The full vertex $W$ in (\ref{vertexW}) is gauge invariant, infra-red finite and 
Bose symmetric. As the vertex is expressed in terms of the function 
${\cal G}$, it is also M\"{o}bius invariant~\cite{bra_vac}. 
Finally, there is no violation of signature conservation:
the incoming three Reggeon state, consisting of one $d$-Reggeon and two $f$-Reggeons,
has even signature; the same holds for the outgoing four Reggeon state (four
$f$-Reggeons).

As a result, the baryonic impact factor introduces a new contribution   
to the Pomeron channel which has no analogue in the photon dipole factor.

\section{Baryon wave functions in the coordinate space}
\label{sec:coord}

The baryon wave function in transverse position space may be easily
obtained by the Fourier transform: 
\beq
\tilde{\Psi}_{\lambda} ^{(\lambda_1,\lambda_2)\lambda_3} 
(\{\alpha_i\},\{\vr_i\},\vP) \; = \;
\int\, {d^2 p_1 \over 2\pi} \,  {d^2 p_2 \over 2\pi} \, {d^2 p_3 \over 2\pi} 
\;\;
\Psi_{\lambda} ^{(\lambda_1,\lambda_2)\lambda_3} 
(\{\alpha_i\},\{\vp_i\},\vP) \; \exp\left(\,i\,\sum_{i=1} ^3 \vp_i \cdot \vr_i 
\right).
\eeq
The result takes a rather simple form:
\beeq
\tilde{\Psi}_{\lambda} ^{(\lambda_1,\lambda_2)\lambda_3} 
(\{\alpha_i\},\{\vr_i\},\vP) 
& = &   
\tilde{\cal N} \; \alpha_1 \alpha_2 \alpha_3 \;\;
\exp\left[-{M^2\over4}\,\sum_i \alpha_i \,(\vr_i-\vR) ^2\, \right]\; 
\exp(\,i\,\vP\cdot\vR\,) \; \times \;
\label{psi_r}
\\
& \times &  \rule{0mm}{10mm} 
\left\{ \rule{0mm}{7mm} 
\;\;\lambda\,M\; 
\delta_{\lambda,\, \lambda_1}\,
\delta_{\lambda,\, -\lambda_2}\,
\delta_{\lambda,\, \lambda_3} \;
[\,\left(\vr_2-\vR\right)\cdot\ve_{\lambda}\,]
\;
[\,\left({\vr_1 - \vr_3} \right)\cdot\ve_{-\lambda}\,]
\right.
\nonumber \\
& + & \rule{0mm}{10mm} 
\hspace{6mm}\lambda\, M\;
\delta_{\lambda,\, -\lambda_1}\,
\delta_{\lambda,\, \lambda_2}\,
\delta_{\lambda,\, \lambda_3} \;
[\,\left(\vr_1-\vR\right)\cdot\ve_{\lambda}\,]
\;
[\,\left(\vr_2-\vr_3\right)\cdot\ve_{-\lambda}\,]
\; + \;
\nonumber \\
& - & \rule{0mm}{10mm} 
\hspace{6mm}
2\,i\;\delta_{\lambda,\,  \lambda_1}\,
\delta_{\lambda,\,  -\lambda_2}\,
\delta_{\lambda,\, -\lambda_3} \; 
\left[\,
(\vr_2  - \vr_3)
\, \cdot\, \ve_{\lambda} \,
\right] 
\; + \;  
\nonumber \\
& - & \rule{0mm}{10mm} 
\hspace{6mm}
\left.
2\,i\;\delta_{\lambda,\,  -\lambda_1}\,
\delta_{\lambda,\,  \lambda_2}\,
\delta_{\lambda,\,-\lambda_3} \; 
\left[
\left( {\vr_1 - \vr_3 } \right)
\cdot\ve_{\lambda} \; 
\right] \;\;
\rule{0mm}{7mm} 
\right\},
\nonumber 
\eeeq
where $\vR$ denotes the light-cone center of mass position vector,
\beq
\vR \,=\, \sum_{i=1}^3 \, \alpha_i\,\vr_i . 
\eeq
The form of the wave function given by Eq.\ (\ref{psi_r}) which follows from 
the Ioffe current shows in detail the angular momentum structure of the baryon 
and the correlations between the angular momenta and quark
helicities. In particular, each scalar product of the type
$\,(\vr_1-\vR)\cdot \ve_{\lambda}\,$ clearly indicates a rotation of quark~1
around the baryon center-of-mass with the orbital angular momentum
$z$-component, $L_z$, equal to $\lambda$. Terms of the type 
$\,(\vr_1-\vr_3)\cdot \ve_{\lambda}\,$ correspond to a similar 
rotation within the quark pair~(1,3). Thus, in the massless quark case,
all components of the baryon carry a non-zero angular momentum $L_z$ 
for the Ioffe operator. An inspection of the momentum space expressions 
(\ref{Thetamass}) shows that for the massive quark case, one may have
Ioffe baryon wave function components with $L_z=0$.

Using Eqs.\ (\ref{bcal}) and (\ref{overlap2}), one may express the baryon impact factors 
${\cal B}_N ^{\,\lambda\lambda'} (\{\vl_{i}\}; \vP,\vP')$
via the overlap function 
${\cal F}^{\;\lambda \lambda'}(\{\vl_{i}\}; \vP,\vP')$
defined in the coordinate space:
\beq
{\cal F}^{\;\lambda \lambda'}(\{\vl_{i}\}; \vP,\vP') \;=\; 
\label{overlap_r}
\eeq
\[
\sum_{\lambda_1,\lambda_2,\lambda_3} \;
\int [d^2 \vr_{i}] \, [d\alpha_{i}] \;\; 
\left[
\tilde\Psi_{\;\lambda'} ^{(\lambda_1,\lambda_2)\,\lambda_3}
\left(\{\alpha_{i}\},\{\vr_{i}\};\vP'\right)   
\right]^* 
\;\,\exp\left(\,-\,i\,\sum_{i=1} ^3 \vl_i \cdot\vr_i \right)\;
\;\,  
\tilde\Psi_{\;\lambda} ^{(\lambda_1,\lambda_2)\,\lambda_3}
\left(\{\alpha_{i}\},\{\vr_{i}\};\vP\right)
. \vspace{3mm}
\] 
It follows from Eqs.\ (\ref{F-function}), (\ref{fnorm}) and~(\ref{overlap_r}) 
that the normalization condition for the wave function reads:
\beq
\sum_{\lambda_1,\lambda_2,\lambda_3} \;
\int [d^2 \vr_{i}] \, [d\alpha_{i}] \;\; 
\left[
\tilde\Psi_{\;\lambda} ^{(\lambda_1,\lambda_2)\,\lambda_3}
\left(\{\alpha_{i}\},\{\vr_{i}\};\vP'\right)   
\right]^* \;
\tilde\Psi_{\;\lambda} ^{(\lambda_1,\lambda_2)\,\lambda_3}
\left(\{\alpha_{i}\},\{\vr_{i}\};\vP\right)
\; = \; 
\delta^{(2)}\left(\,\vP\, - \,\vP'\, \right).
\label{wavenorm}
\eeq
It is instructive to evaluate a contribution to the baryon two-gluon impact 
factor $[\delta{\cal B} _{2;0}]^{\{1,2\}}$ corresponding to a dipole-like piece, 
e.g.\ to $D_{2;0} ^{\{1,2\}}$, in the coordinate representation.
The gluon color labels are $a_1$ and $a_2$ and momenta are denoted 
by $\vk_1$ and $\vk_2$ respectively.
One obtains: 
\beq
\label{delb}
\left[\delta{\cal B}_{2;0} ^{\,\lambda\lambda'} (\{\vl_{i}\}; \vP,\vP')\right]^{\{1,2\}}
\; = \; {1\over 2}\;(-ig)^2\,{\delta^{a_1 a_2} \over 2N_c}\, \times
\eeq
\[
\times \;
\sum_{\lambda_1,\lambda_2,\lambda_3} \;
\int [d^2 \vr_{i}] \, [d\alpha_{i}] \;\; 
\left[
\tilde\Psi_{\;\lambda'} ^{(\lambda_1,\lambda_2)\,\lambda_3}
\left(\{\alpha_{i}\},\{\vr_{i}\};\vP'\right)   
\right]^*
\;\,
\tilde\Psi_{\;\lambda} ^{(\lambda_1,\lambda_2)\,\lambda_3}
\left(\{\alpha_{i}\},\{\vr_{i}\};\vP\right)
 \; \times 
\] 
\[
\times \;
\left[
e^{-\,i\,(\vk_1+\vk_2) \cdot\vr_1 \, }
\; + \;
e^{-\,i\,(\vk_1+\vk_2) \cdot\vr_2 \, }
\;- \;
e^{-\,i\,\vk_1\cdot\vr_1  \,-\, i\,\vk_2\cdot\vr_2  \,}
\; - \;
e^{-\,i\,\vk_1\cdot\vr_2  \,-\, i\,\vk_2\cdot\vr_1  \,}\;
\right].
\]
Assuming, for simplicity, the forward kinematics, $\vk_1 = \vk = -\vk_2$,
one may rewrite the eikonal factors in the last line of (\ref{delb}) in 
a factorized form, found in the case of the color dipole scattering,  
\beq
\left[\,1\,-\,e ^{i\vk\cdot(\vr_2-\vr_1)} \right]\; 
\left[\,1\,-\,e ^{i\vk\cdot(\vr_2-\vr_1)} \right]^*\, .
\eeq
This equivalence of the structures holds also beyond the forward limit (note that, 
for nonzero $\vP$, $\vP'$ the wave functions $\tilde\Psi_{\;\lambda'} 
^{(\lambda_1,\lambda_2)\,\lambda_3}$ contain the phase factors 
$e^{i\vP \vR}$).
In Eq.\ (\ref{delb}), the prefactor 1/2 in the first line reflects 
the relative weight between the color dipole scattering amplitude 
and the scattering amplitude of the dipole-like components of the baryon.

\section{The quark--diquark limit}

In many phenomenological applications the nucleon is represented as
a bound state of quark and a tightly bound diquark. The transverse size of the 
diquark is then assumed to be much smaller than the size of the baryon, and
the diquark state emerges in an anti-triplet color representation. 
In this approximation the baryon should resemble an (asymmetric) color dipole. 
It is interesting to
analyze the properties of our baryon impact factor in this limit.
Formally, the quark--diquark limit corresponds to the limit where 
the transverse separation of two quark lines shrinks to zero, and 
a $t$-channel gluon no longer distinguishes between the two quark lines. 
In momentum space, as seen in (\ref{overlap_r}), the overlap function 
then only depends upon the sum of the momenta of all gluons coupled to the two 
coinciding quark lines.
To be definite, let us assume that quarks 2~and~3 move close to each other.
Then all overlap functions $F$ degenerate to a function $F^{1(23)}$ with only 
two arguments:
\beq
F(\vk_1,\vk_2,\vk_3) \;\; \underset{3\to 2}{\longrightarrow} \;\; 
F^{1(23)} (\vk_1,\vk_2 + \vk_3).
\eeq
(note that the limit $F^{1(23)}(\vk_1,\vk_2)$ is not necessarily 
symmetric in its arguments). Applying this argument to the three dipole-like 
terms in (\ref{dipole}) we immediately see that the dipole-like component 
$D_{2;0}^{\{2,3\}}$ vanishes if lines $2$ and $3$ are contracted: 
this is the well-known limit of a dipole with vanishing size 
(color transparency). In more detail, (\ref{D_2,23}) shows that all terms 
in this impact factor tend to $F^{1(23)}(0,\vk_1+\vk_2)$, and
they cancel due to opposite signs. 
The remaining dipole-like components $D_{2;0}^{\{1,2\}}$ and 
$D_{2;0}^{\{1,3\}}$ become equal: 
\beq
D_{2;0}^{\{1,2\}}(\vk_1,\vk_2), \; D_{2;0} ^{\{1,3\}}(\vk_1,\vk_2)\;\; 
\underset{3\to 2}{\longrightarrow} \;\; D_{2;0} ^{\{1,(23)\}}(\vk_1,\vk_2),
\eeq
with
\beq
D_{2;0}^{\{1,(23)\}}(\vk_1,\vk_2) \; = \; {-g^2 \over 12} \, 
\left[
F^{1(23)}(0,\vk_1+\vk_2) +  
F^{1(23)}(\vk_1+\vk_2,0) 
-F^{1(23)}(\vk_1,\vk_2)
-F^{1(23)}(\vk_2,\vk_1)
\right].
\eeq
As we already discussed at the end of Section~4.2, in (\ref{dipole}) each 
dipole-like term carries a color factor $1/2$, compared to a genuine 
color dipole factor seen in a color singlet quark-antiquark system. 
Since in the quark--diquark limit $D_{2;0}^{\{2,3\}}$ 
vanishes and the contributions from  $D_{2;0}^{\{1,2\}}$ and  $D_{2;0}^{\{1,3\}}$ 
coincide, this part of the baryonic impact factor 
adds up to a standard dipole contribution 
$D_{2;0}(\vk_1,\vk_2) =  2D_{2;0}^{\{1,(23)\}}(\vk_1,\vk_2)$.
 
Next, we turn to the three-gluon impact factors.
In the Pomeron channel, one finds only reggeizing pieces of the 
quark--diquark dipole impact factor. In the odderon channel, the
function $E(\vk_1,\vk_2,\vk_3)$ degenerates to the structure
found in the $\,\gamma^* \to \eta_c\,$ transition impact factor,
which couples only to the Bartels-Lipatov-Vacca (BLV) odderon~\cite{blv2} 
but not to the Janik-Wosiek solution~\cite{jan_wos}:
\beq
E_{3;0}(\vk_1,\vk_2,\vk_3) \;\; \underset{3\to 2}{\longrightarrow} \;\; 
E_{3;0}^{\{1,(23)\}} (\vk_1,\vk_2,\vk_3),
\eeq
with 
\[
E_{3;0}^{\{1,(23)\}} (\vk_1,\vk_2,\vk_3) \;  =  \; {ig^3 \over 12}\;
\left[
F^{1(23)} (\vk_1,\vk_2+\vk_3) \,-\, F^{1(23)} (\vk_2+\vk_3,\vk_1) \,+\, 
\right.
\]
\[
+\,F^{1(23)} (\vk_2,\vk_1+\vk_3) \,-\, F^{1(23)} (\vk_1+\vk_3,\vk_2) \, 
+ \; F^{1(23)} (\vk_3,\vk_1+\vk_2)\,-\, F^{1(23)} (\vk_1+\vk_2,\vk_3) \, +
\]
\beq
\left.
+ \, F^{1(23)} (\vk_1+\vk_2+\vk_3,0) \,-\, F^{1(23)} (0,\vk_1+\vk_2+\vk_3)
\right].
\eeq 

For the four gluon case, one finds the standard reggeizing 
pattern of $D_{2;0}^{\{1,(23)\}}$ and of $E_{3;0}^{\{1,(23)\}}$ in the
Pomeron and the odderon channel, respectively. 
The structure $Q_{4;0}$ vanishes in the quark--diquark limit. This is the 
result of a nontrivial cancellation of all three lines of Eq.~(\ref{eq:q40}),
making use of the identity (\ref{eq:deltad}). 
The pattern given by the impact factors in the small diquark limit is preserved 
by the small~$x$ evolution, in particular $Q_4$ vanishes.

In summary, we have verified that, in the quark--diquark limit, the baryon 
reduces to a dipole-like object with an asymmetric wave function, 
as it was expected. Conversely, our analysis shows that, outside the 
diquark limit, the baryon impact factor contains a new piece (related to 
$Q_{4;0}$) which accompanies the appearance of the third dipole-like 
term, $D_{2;0}^{\{2,3\}}$. A more detailed study of the question, to 
what extent the baryon wave functions actually favors a diquark state, should 
start from the Fourier transform of the overlap function, (\ref{overlap_r}), 
which describes the distribution of the quarks in transverse coordinate space.
Further work along these lines is in progress.  

\section{Discussion}

In this paper we have investigated the high energy behavior of a baryonic 
state. We have studied the structure of a baryonic impact factor,  
its coupling to multi-gluon exchanges and the rapidity evolution of 
the $t$-channel gluon states. We found it convenient to follow 
very much the same approach, which has been developed and used for the 
high energy behavior of a virtual photon (or a heavy quarkonium state). 
For the scattering of such mesonic states, in the leading logarithmic 
approximation and in the large-$N_c$ limit, the high energy behavior 
allows for the interpretation in terms of color dipoles, 
and one of the motivations of our investigation was the question to what 
extent this attractive physical picture can be used also for the scattering 
of baryonic states.

Compared to the quark-antiquark system created by the photon 
(or a heavy vector meson), the high energy scattering of baryonic systems 
consisting of three quarks shows similarities, but also striking differences. 
First, there is a component of the baryonic impact factor 
in which two of the three quarks interact with the target 
whereas the third one acts as a spectator. Here the two-quark subsystem 
behaves very much in the same way as the color singlet dipole of the
quark-antiquark system. In particular, the rapidity evolution is the 
same as in the case of a virtual photon. This configuration, however,  
extends beyond the picture of a small ``diquark state'': we have shown that, 
in the diquark limit, we recover the dipole picture. But the 
spectator quark is not necessarily linked (in transverse space) to one of the 
participating quarks, and our analysis includes also this more general 
configuration. Second, there is the piece of the baryon impact factor to which 
the $C$-odd three gluon state (odderon) couples. 
Third, a new piece of the baryonic impact factor exists which couples to 
a $C$-even three gluon $t$-channel state, and there is a new vertex which 
describes the transition from this three gluon state to the four gluon 
(two Pomeron) state. In the quark-antiquark case, there is no analogue 
of this contribution.

This third piece may actually be quite essential for the restauration 
of $s$-channel unitarity in baryon scattering and can therefore not be 
neglected. Namely, let us consider the scattering of a hypothetical heavy baryon on a large 
nuclear target; this represents the analogue of the Balitsky-Kovchegov problem 
for the color dipole scattering. Based on our results, the baryon scattering amplitude ${{\cal B}}$
can be written symbolically as a sum of the following pieces:
\beq
\label{barsymb}
{\,{\cal B}} \;\; = \;\;  
\overbrace{{{\cal D}_2 ^{\{1,2\}}} \,+\,
{{\cal D}_2 ^{\{1,3\}}} \,+\,
{{\cal D}_2 ^{\{2,3\}}} \,+\,
{{\cal Q}_4} }^{C\mathrm{-even}} \;+\;
\overbrace{{\,{\cal E}_3}}^{C\mathrm{-odd}}.
\eeq
Here the first three terms, $\,{{\cal D}_2 ^{\{i,j\}}}\,$,  stand for the 
dipole-like contributions in which the baryon couples to the same two-point gluon 
correlator as the color dipole in the scattering of a virtual photon. 
The strength of this coupling, however, 
is only $1/2$ of that for the photon dipole. The pieces  $\,{{\cal Q}_4}\,$ 
and $\,{\,{\cal E}_3}\,$ probe
three-point gluon correlators: the $C$-even and $C$-odd ones respectively.  
As it was observed in the case of the color dipole in deep inelastic scattering, 
where only a single BFKL 
Pomeron could couple to the dipole, we again see no indications of a direct 
two Pomeron coupling to the valence degrees of freedom of the baryon. 
If we assume that the two-gluon distribution probed by the first three terms in 
Eq.\ (\ref{barsymb}) is consistent with saturation of the black disc limit for color dipoles of the sizes 
given by the baryon geometry, then the $T$-matrices for each of the
$\,{{\cal D}_2 ^{\{i,j\}}}\,$ components would tend to $1/2$, and 
the total contribution of the dipole-like pieces to the baryon $T$-matrix 
would amount to~$3/2$. This would mean that $s$-channel unitarity can be 
maintained only if $\,{{\cal Q}_4}\,$ and $\,{\,{\cal E}_3}\,$ give a 
combined contribution to the $T$-matrix smaller than $-1/2$.
Thus, the three-Reggeon states  $\,{{\cal Q}_4}\,$ and 
$\,{\,{\cal E}_3}\,$ seem to be essential to guarantee the
$s$-channel unitarity. Interestingly enough, one might go even further and 
arrive at a quantitative prediction:  if one postulates that
the $T$-matrices, both for the color dipole and the baryon scattering 
at very large energies saturate the unitarity limit --- one then finds 
that in the black disc limit:\,
(i) the $C$-odd three point gluon correlator should vanish; this comes 
from the requirement that both proton and anti-proton scattering should reach
the black disc limit, despite the fact that the amplitude 
$\,{\,{\cal E}_3}\,$ has opposite signs in these two cases;\,
(ii) the $C$-even three point correlator is strongly constrained: 
when coupled to the impact factor $Q_{4;0}$ it must lead to the
scattering amplitude equal to $-1/2$. In the diquark limit, both 
$D_2^{\{2,3\}}$ and ${\cal Q}_4$ vanish, and unitarization proceeds 
in the same way as in the dipole case.

We interpret these results as a strong indication that, in the 
context of baryon scattering, QCD Reggeon field theory has to be extended 
beyond the theory of BFKL Pomerons and their interactions. 
First, it is difficult to justify the large-$N_c$ limit, which, 
in the scattering of virtual photon and mesonic states, 
allows to reduce the evolution of BKP states consisting of $2n$-gluon to 
the propagation of $n$~BFKL Pomerons. 
Second, the three gluon state (and its BKP evolution) seems 
to play an important r\^{o}le, not only in the odderon channel. 
As we have pointed out, this phenomenon is closely connected with the 
existence of the $d$ Reggeon, the even signature partner of the 
(odd signature) reggeized gluon. 

On a deeper level one may speculate that there exists an intimate
connection between the number of valence objects in the impact factor
in the fundamental $SU(N_c)$ representation and the maximal number of 
Reggeons in the BKP state which couple to the impact factor. 
For the quark-antiquark color dipole only the two-Reggeon 
BFKL Pomeron couples, and for the baryon containing three quarks we
have both two- and three-Reggeon states. We may conjecture that the
number of the different BKP states that couple to the baryon in  
$SU(N_c)$ gauge theory is related to the number of Casimir operators of
the gauge group. There exist two Casimir operators of the $SU(3)$ gauge 
group, and QCD Reggeon field theory (whose basic degrees of freedom are the 
reggeized gluons) exhibits two `fundamental excitations' which, in the 
leading-log approximation, are represented by the two-gluon BFKL Pomeron 
and by the three-gluon odderon state. 
For a high energy $SU(N_c)$ baryon we expect that the impact factor, 
consisting of $N_c$ quarks in the fundamental representation, 
would exhibit all the $2$, $3$, \ldots, $N_c$ gluon states, and
it would hint that the number of fundamental glue excitations may be related
to the $N_c-1$ Casimir operators of $SU(N_c)$.    
It seems natural that the gauge group invariants should be mapped 
onto gauge invariant BKP states. The explicit connection, however, 
has not been yet established.

Turning to more practical and phenomenological applications, 
in this paper we have considered a baryonic state consisting of three 
massive quarks being in a proton-like configuration. One can view 
such a `heavy baryonium' state as a convenient theoretical laboratory, 
very much in the same spirit as previous work on high energy QCD
has made use of `heavy onium' states. On the other hand, we feel that 
our results might also allow for immediate phenomenological applications.
In particular, we have proposed a relativistic invariant model of the proton
wave function, including the helicity structure and correlations between
helicities and quark angular momenta. Both the model itself and the 
calculational technique applied may be useful in studies of polarized 
scattering of the proton and of the proton form-factors.
Another potential place of interest is the intermediate $t$~region of 
proton--proton elastic scattering where, in the days of ISR experiments, 
a very simple three gluon model had a striking phenomenological 
success~\cite{landshoff}. It should also be
quite interesting to study other applications of the model in the context
of elastic $pp$ and $p\bar p$ scattering and exclusive diffraction at 
RHIC, Tevatron and the LHC. Finally, we would like to view our study
as a preparation for a QCD analysis of multiple scattering in $pp$ 
collision at the LHC.

\section*{Acknowledgments}
We especially acknowledge the help of G.P.~Vacca who contributed in the 
early stage of this work. 
We thank C.~Ewerz, L.~Lipatov, and M.~Salvadore for interesting  discussions,
and A.~Bia\l{}as, S.~Bondarenko, M.~Diehl, 
Yu.~Kovchegov, E.~Levin and R.~Peschanski for useful comments.
We thank the Galileo Galilei Institute in Florence for the support
in the initial phase of this research project.
L.M.\ gratefully acknowledges the support of the DFG grant SFB~676 and
the grant of the Polish State Committee for Scientific Research 
No.~1~P03B~028~28.

\appendix

\section{Appendix}
\subsection{Spinorial matrix elements}

The calculations of the baryon wave functions and of the 
baryon scattering amplitudes are performed using the 
light-cone formalism summarized in \cite{bro_lep}.

Thus we employ the spinor basis defined by
\beq
\left.
\begin{array}{c}
u_{\uparrow}(p) \\
u_{\downarrow}(p)   
\end{array}
\right\} = {1\over \sqrt{p^+}} (p^+ + \hat\beta m + \hat{\pmb\alpha} \cdot \vp)
\times 
\left\{
\begin{array}{c}
\chi(\uparrow) \\
\chi(\downarrow)
\end{array}
\right.
\eeq
and
\beq
\left.
\begin{array}{c}
v_{\uparrow}(p) \\
v_{\downarrow}(p)   
\end{array}
\right\} = {1\over \sqrt{p^+}} (p^+ - \hat\beta m + \hat{\pmb\alpha} \cdot \vp)
\times 
\left\{
\begin{array}{c}
\chi(\downarrow) \\
\chi(\uparrow),
\end{array}
\right.
\eeq
where
\beq
\chi(\uparrow) \;=\; 
{1\over \sqrt{2}}\,
\left( \begin{array}{c} 1 \\ 0 \\1 \\0 \end{array} \right)\, , \qquad
\chi(\downarrow) \;=\; 
{1\over \sqrt{2}}\,
\left( \begin{array}{c} 0 \\ 1 \\ 0 \\-1 \end{array} \right)\, 
\eeq
in the Dirac representation, and the Dirac matrices 
$\hat\beta$ and $\hat\alpha$ are related to the $\gamma$-matrices through   
$\hat\beta = \gamma^0$ and $\hat{\alpha}^{s} = 
\gamma^0\gamma^{s}$; $m$ is the mass of a fermion
(or an anti-fermion).
In the infinite momentum frame, when $p^+ \to \infty$ 
these spinors tend to the helicity eigenstates,
$u_{\uparrow\downarrow}(p) \to u_{\pm}(p)$, 
$v_{\uparrow\downarrow}(p) \to v_{\pm}(p)$. 

In the calculation of the baryon $\to$ quarks transition amplitudes
it is sufficient to employ spinor matrix elements given in the following 
tables. Note that we consider a general case in which the masses of the 
spinors $u$ (or $v$) and $u'$ are given by $m$ and $m'$, respectively.
\begin{equation}
\begin{array}{c|c|c}
\parbox{5cm}{\begin{center}
\small Matrix element \\
$\bar u'_{\lambda'}(p) \ldots u_{\lambda}(q)$
\end{center}}  
&
\parbox{5cm}{\begin{center}
\tiny 
{\small $\lambda \, \to \lambda'$} \\
$\uparrow \;\to\;\uparrow$ \\
$ \downarrow\;\to\;\downarrow $\end{center}}
&
\parbox{5cm}{\begin{center}\tiny 
{\small $\lambda \, \to \lambda'$}\\
$\uparrow\;\to\;\downarrow$ \\
$\downarrow\;\to\;\uparrow $\end{center}} 
\\ \hline
{\bar u' (p) \over \sqrt{p^+}}\, \gamma^+ \,{u (q) \over \sqrt{q^+}} &
2 & 0 \rule{0cm}{10mm}
\\
{\bar u' (p) \over \sqrt{p^+}}\, \gamma^- \,{u (q) \over \sqrt{q^+}} 
&
{2\over p^+q^+}[(\vp \cdot \ve_{\mp}) \, (\vq \cdot \ve_{\pm}) \; + \; m m'] 
&
\mp{2\over p^+q^+}(m\;\vp \cdot \ve_{\pm} \, -\,  m'\;\vq \cdot \ve_{\pm})
\rule{0cm}{10mm}
\\
{\bar u' (p) \over \sqrt{p^+}}\, \gamma_{\perp}^s \,{u (q) \over \sqrt{q^+}} 
&
\eta^s_{\pm}\, {\vp \cdot \ve_{\mp} \over p^+} \;+\; 
\eta^s_{\mp}\, {\vq \cdot \ve_{\pm} \over q^+} 
&
\pm \eta^s _{\pm}\left( {m' \over p^+} \,-\, {m \over q^+} \right) 
\rule{0cm}{10mm}
\\ & & 
\rule{0cm}{5mm}
\\
\hline 
\end{array}\nonumber
\end{equation}

\begin{equation}
\begin{array}{c|c|c}
\parbox{5cm}{\begin{center}
\small Matrix element \\
$\bar v'_{\lambda'}(p) \ldots u_{\lambda}(q)$
\end{center}}  
&
\parbox{5cm}{\begin{center}
\tiny 
{\small $\lambda \, \to \lambda'$} \\
$\uparrow \;\to\;\uparrow$ \\
         $ \downarrow\;\to\;\downarrow $\end{center}}
&
\parbox{5cm}{\begin{center}\tiny 
{\small $\lambda \, \to \lambda'$} \\
$\uparrow\;\to\;\downarrow$ \\
$\downarrow\;\to\;\uparrow $\end{center}} 
\\ \hline
{\bar v' (p) \over \sqrt{p^+}}\, \gamma^+ \,{u (q) \over \sqrt{q^+}} &
0 & 2 \rule{0cm}{10mm}
\\
{\bar v' (p) \over \sqrt{p^+}}\, \gamma^- \,{u (q) \over \sqrt{q^+}} 
&
\mp{2\over p^+q^+}(m\;\vp \cdot \ve_{\pm} \, + \,  m'\;\vq \cdot \ve_{\pm})
&
{2\over p^+q^+}[(\vp \cdot \ve_{\mp}) \, (\vq \cdot \ve_{\pm}) \; - \; m m'] 
\rule{0cm}{10mm}
\\
{\bar v' (p) \over \sqrt{p^+}}\, \gamma_{\perp}^s \,{u (q) \over \sqrt{q^+}} 
&
\mp \eta^s _{\pm}\left( {m' \over p^+} \,+\, {m \over q^+} \right) 
&
\eta^s_{\pm}\, {\vp \cdot \ve_{\mp} \over p^+} \;+\; 
\eta^s_{\mp}\, {\vq \cdot \ve_{\pm} \over q^+} 
\rule{0cm}{10mm}
\\ & & 
\rule{0cm}{5mm}
\\
\hline 
\end{array}\nonumber
\end{equation}

As an example, we apply the above formulae to evaluate 
\beeq
{\left[ \, \bar d _{\lambda_3}(p_3) \, \gamma_\mu\, w_{\lambda}(P) \, \right] 
\,\cdot \, \left[ \, \bar u _{\lambda_1}(p_1)\,\gamma^\mu\, v_{\lambda_2}(p_2) \,\right] 
\over \sqrt{P^+ \,p_1^+\, p_2^+\, p_3^+}}
& = &
{1\over 2} \,
{\left[ \, \bar d _{\lambda}(p_3) \, \gamma^+\, w_{\lambda}(P) \, \right] 
\,\cdot \, 
\left[ \, \bar u _{\lambda_1}(p_1) \, \gamma^-\, v_{\lambda_2}(p_2)\,\right] 
\over
\sqrt{P^+ \,p_1^+\, p_2^+\, p_3^+}} \nonumber\\
& + &  
{1\over 2} \,
{\left[ \, \bar d _{\lambda}(p_3) \, \gamma^-\, w_{\lambda}(P) \, \right] 
\,\cdot \, 
\left[ \, \bar u _{\lambda_1}(p_1) \, \gamma^+\, v_{\lambda_2}(p_2)\,\right] 
\over
\sqrt{P^+ \,p_1^+\, p_2^+\, p_3^+}} \nonumber\\
& - &
{\left[ \, \bar d _{\lambda}(p_3) \, \gamma_{\perp}^s\, w_{\lambda}(P) \, \right] 
\,\cdot \, 
\left[ \, \bar u _{\lambda_1}(p_1) \, \gamma_{\perp}^s\, v_{\lambda_2}(p_2)\,\right] 
\over
\sqrt{P^+ \,p_1^+\, p_2^+\, p_3^+}} 
\label{bl_example}
\eeeq
for $\lambda=\lambda_1=-\lambda_2=\lambda_3 = +1$. 
The prefactors: $1/2$, $1/2$ and $-1$ on the 
r.h.s.\ are the only non-vanishing elements of the covariant 
metric tensor $g_{\mu\nu}$ in the light-cone coordinates.
In the calculations we find it useful to make use of the following identities 
for transverse complex vectors $\ve_{\pm}$: $\ve_+ ^* = \ve ^-$, $\;\ve_{\pm}^2 = 0$,
$\; \ve_{\pm}\cdot \ve_{\mp} = |\ve_{\pm}|^2 = 2$.
Thus, assuming that the light quark masses vanish, we obtain:
\beeq
{\left[ \, \bar d _{\lambda_3}(p_3) \, \gamma_\mu\, w_{\lambda}(P) \, \right] 
\,\cdot \, 
\left[\, \bar v _{\lambda_2}(p_2)\,\gamma^\mu\, u_{\lambda_1}(p_1) \,\right]^* 
\over \sqrt{P^+ \,p_1^+\, p_2^+\, p_3^+}}
& = & {2[(\vp_2 \cdot \ve_{-})(\vp_1 \cdot \ve_{+}) ]^* \over p_1^+ p_2^+}
+ {2(\vp_3 \cdot \ve_{-})(\vP \cdot \ve_{+}) \over P^+ p_3^+}
\nonumber \\
& & 
-{2(\vp_3\cdot\ve_{-})(\vp_2\cdot\ve_{-})^* \over p_2^+ p_3^+}
-{2(\vP\cdot\ve_{+})(\vp_1\cdot\ve_{+})^* \over P^+ p_1^+} \nonumber \\
& = &  
2\,\left[{\vp_2\cdot\ve_{+} \over p_2^+} - {\vP\cdot\ve_{+} \over P^+}\right]
\left[{\vp_1\cdot\ve_{-} \over p_1^+} - {\vp_3\cdot\ve_{-} \over p_3^+}\right].
\eeeq
Using an identity\footnote{For a non-zero quark mass $m$, the relation 
holds approximately in the large energy limit, 
$\;\bar d_{\lambda_3}(p_3)\, \gamma_5 \; = \; 
\lambda_3 \, \bar d_{\lambda_3}\,(p_3)  +  O(m/p_3^+)$.}
 $\;\bar d_{\lambda_3}(p_3) \, \gamma_5 \; = \; 
\lambda_3 \, \bar d_{\lambda_3}(p_3)\,$, and
relation (\ref{etaprod}),
one obtains one of the matrix elements described by (\ref{Theta1}).
The matrix elements for all remaining choices of helicities can be 
derived in the same way.

\subsection{A reduction formula for spinors in high energy limit}

We shall prove the following identity for massive Dirac spinors:
\beq
\label{shift0}
\bar u(p)\,\hat q \, (\hat p + m + \hat k) \; = \; 
2\, p\cdot q \; \bar u(p+k) + \ldots,
\eeq
which holds, at the leading accuracy in  $s \simeq 2 p\cdot q$,
in the high energy limit: $s \gg q^2,\, k^2,\,  m^2,\, p\cdot k,\, 
q\cdot k\;$ etc., and for $k_{\perp} \gg k^+, k^-$.
This identity is a useful tool for deriving quark scattering 
amplitudes by multi-gluon couplings in the eikonal approximation.
Using the spinor equation of motion, $\,\bar u(p) (\hat p - m) = 0\,$, 
we get
\beq
\bar u(p)\,\hat q \,(\hat p + m + \hat k) \;= \;
\bar u(p)\, \left(\,2p\cdot q \, + \, \hat q \, \hat k\,\right) \;\simeq\; 
s\; \bar u(p) 
\left(\,1 + {1\over 2s}\left[\,\hat q\,,\,\hat k\,\right]\,\right),
\eeq
where we used the fact that the anticommutator 
$\{\,\hat q\,,\,\hat k\,\} = 2k \cdot q \ll s$. 
Furthermore, using the light-cone variables, as defined in Sec.\ 3,
we have
\beq
\left[\,\hat q\,,\,\hat k\,\right] \;=\; 
-2i\,\hat\sigma_{\alpha\beta}\, q^\alpha\,k^\beta \;\simeq\; 
-2i\,\hat\sigma_{-\,r}\, q^-\; k_{\perp}^{\,r},
\eeq 
where $r$ is the Lorentz index of the transverse coordinates.
Thus one obtains
\beq
\bar u(p)\,\hat q \,(\hat p + m + \hat k) 
\;\simeq\; 
s\; \bar u(p) 
\left(\,1 - {i\,\hat\sigma_{-\,r}\; k_{\perp}^{\,r} \over p^+}\,\right).
\label{shift1}
\eeq
The matrices $\hat\sigma^{\alpha\beta}$ are proportional to the generators of 
the Lorentz transformations of the Dirac spinors:
\beq
\exp \left( -{i\over 4} \sigma_{\alpha \beta} 
\omega^{\alpha\beta} \right)\, u(p)
=  u(\Lambda(\omega) p), \qquad 
 \bar u(p)\,\exp \left( {i\over 4} \sigma_{\alpha \beta} 
\omega^{\alpha\beta} \right)
= \bar u(\Lambda(\omega) p),
\label{shift2}
\eeq
where 
\beq
(\Lambda(\omega) p)^\mu = [\Lambda(\omega)] ^{\mu} _{\;\;\nu} p^\nu, \qquad
\Lambda(\omega)  = \exp\left( {1\over 2} \omega^{\alpha\beta} 
L_{\alpha\beta} \right),
\eeq
and the generators of Lorentz transformations in the vector representation
read
\beq
[L_{\alpha\beta}]^{\mu} _{\;\;\nu}\; =\; 
g^{\mu} _{\;\;\alpha} \,g_{\nu \beta} \,-\,g^{\mu} _{\;\;\beta}\, g_{\nu \alpha}\,.
\eeq
Since the parameter multiplying $\hat\sigma_{\,-\,r}$ in 
Eq.\ (\ref{shift1}) is small, 
$\beta^r = k_{\perp} ^r / p^+ \,\ll\, 1$, one may write 
\beq
\bar u(p)\,\left(\,1 - i\,\hat\sigma_{\,-r}\; \beta^r \,\right) 
\; = \;
\bar u(p)\, \left[\exp \left(\, -i\,\hat\sigma_{\,0r}\beta^r / 2 
\right) \,
\exp\left(\, i\,\hat\sigma_{\,3r}\beta^r/2 \,\right)\,\right] 
\; + \; O(\beta^2), 
\label{shift3}
\eeq 
where we used the identity $\gamma_- = {1\over 2}(\gamma_0 - \gamma_3)$.
This equation corresponds to two subsequent infinitesimal Lorentz 
transformations acting on $\bar u(p)$ with the parameters 
$\omega_1^{r0} = -\omega_1 ^{0r} = \beta^r$ and
$\omega_2^{3r} = -\omega_2 ^{r3} = \beta^r$ 
(and all other components  $\omega_{1,2} ^{\alpha\beta} = 0$).
This is an infinitesimal boost along the transverse direction $\pmb\beta$, and 
an infinitesimal rotation in the plane spanned by the transverse 
vector $\pmb\beta$ around the $z$-axis. 
Using (\ref{shift2}) one sees that, in leading order in $\beta^r$,  
the boost transforms $p$ in the following way $\;:p^0 \to p^0,\; 
\vp \to \vp + p^0 \pmb\beta,\; p^3 \to p^3,\,$ and the rotation acts as:
$\;p^0 \to p^0,\; \vp \to \vp + p^3 \pmb\beta, \; p^3 \to p^3$.   
Thus one obtains
\beq
\bar u(p)\,\left(\,1 - i\,\hat\sigma_{\,-r}\; \beta^r \,\right) 
\; = \; \bar u(p') \; + \; O(\beta^2),
\eeq
with $p' = (p^0,\vp + {\pmb\beta}p^+, p^3)$. 
This proves Eq.~(\ref{shift0}).
The equation for multiple eikonal couplings, Eq.~(\ref{multishift}), follows
immediately from Eq.~(\ref{shift0}), after all spinor contractions, 
$\;\hat q \,(\hat p - \hat k_1 - \ldots -\hat k_i + m )\, \hat q\; \simeq 
\; 2p\cdot q \; \hat q,\,$ are executed.

\end{document}